\documentclass[12pt]{article}

\usepackage[english]{babel}

\usepackage[a4paper,margin=2cm,marginparwidth=1.75cm]{geometry}

\usepackage{ragged2e} 

\usepackage{amsmath}
\usepackage{graphicx}
\usepackage[colorlinks=true, allcolors=red]{hyperref}
\usepackage[super]{natbib}
\usepackage{multirow}
\usepackage{booktabs}
\usepackage{caption}
\usepackage{amsmath,amssymb}
\usepackage{rotating}
\usepackage{placeins}
\usepackage{subcaption}

\title{Estimating marginal association between paired outcomes in clustered data with an application to analyzing caries and periodontal disease data from the  National Health and Nutrition Examination Survey}
\author{Owen Visser, Steven Levy, Bruce Dye,\\ Shaoping Zhang, Victor Agboli, Somnath Datta}
\date{\today}

\begin{document}
\justifying
\maketitle

\begin{abstract}
The within-cluster subgroup composition of units can inform on the marginal distribution of those units outcomes, denoted informative subgroup size (ISS). Moreover, within-cluster subgroup composition of paired outcomes can allow one outcome to inform upon the marginal distribution of the other. If this type of ISS ignored, it can bias marginal association estimates by allowing subgroup composition to influence how clusters contribute to the analysis. We develop two weighting approaches for estimating marginal association between paired outcomes in this setting. The weights are derived using within-cluster resampling arguments and extend conventional cluster weighting to paired response subgroups. We adapt an existing test of marginal dependence to assess ISS by examining this subgroup structure in both directions. In simulated data exhibiting this bidirectional ISS, pair-based weighting reduced bias when association was primarily carried through the paired observations. These weights also allow scientifically or clinically meaningful dichotomizations to enter directly into the weighting scheme, with the choice of subgroup definition further improving estimation in several simulated settings. Application to NHANES oral-health data showed strong evidence of ISS for filled-surface outcomes across subgroups defined by clinical attachment loss and pocket depth, with the corresponding periodontal--caries associations generally small and positive. These methods provide a practical approach for estimating marginal association when paired subgroup composition is informative.
\end{abstract}

\section{Introduction}\label{intro}

Informative cluster size (ICS) occurs when the number of observational units within a cluster is associated with the outcome under study\cite{hoffman2001within}. Modern literature on ICS developed from within-cluster resampling (WCR), which gives clusters equal contribution regardless of their observed size\cite{hoffman2001within}. Proceeding research showed that inverse cluster-size weighting is asymptotically equivalent to WCR, providing a computationally simpler route to the same marginal target\cite{williamson2003marginal}. Hence, their formulation of cluster-weighted generalized estimating equations (CWGEEs), where observations are weighted by the inverse of their cluster size. Subsequent work on clustered longitudinal data further showed that ordinary generalized estimating equations (GEEs) may be biased under ICS, while WCR and cluster weighting recover the desired cluster-based marginal interpretation\cite{wang2011inference}.

It was later shown that CWGEEs may still fail when the number of observations sharing a particular within-cluster covariate value is itself informative; we denote this as informative subgroup size (ISS). Thus, the doubly weighted generalized estimating equations (DWGEEs) was introduced to account for both ICS and ISS\cite{huang2011informative}. This idea was extended to distributional comparisons when the number of subjects belonging to a group within a cluster is informative\cite{dutta2016rank}. More recent work has placed ISS within a broader multi-level inferential framework, emphasizing that the appropriate marginalization depends on which strata level the informativeness enters the clustered data structure\cite{anyaso2025can}.

Related works considered marginal association itself and showed how inverse cluster-size weighting can be incorporated through estimating equations to reduce bias from ICS\cite{lorenz2011marginal}. Later, this work was extended to paired and unpaired clustered data, providing a general basis for multivariate association targets\cite{lorenz2018inferring}. These results provide the machinery needed to estimate marginal association once an appropriate weighting scheme has been specified.

Hence, the remaining gap is primarily one of weighting. Existing work provides weights for ICS and for ISS in marginal means, regression, and distributional comparisons, but not for estimating association between two paired outcomes where the effect of ISS on either outcome is driven jointly\cite{andersen2003generalised}. 

The ICS and ISS issues arise naturally in dental research, where the number of teeth remaining in a patient's mouth may reflect oral health status\cite{huang2011informative, wang2011inference}. Moreover, this informativeness may also occur within the composition of a patients teeth. For example, a mouth containing a greater proportion of caries disease may systematically differ from one containing periodontal disease. Caries and periodontal disease are biologically and clinically related, although previous analyses generally suggest only a small positive association between them\cite{durand2019dental}. Additionally, the observed dentition reflects prior disease, treatment, and tooth loss, so both the total number of observed teeth and their distribution across disease states may be informative. Recent work by Anyaso-Samuel and Datta showed that the number of teeth associated with particular levels of a continuous periodontal covariate may remain informative even after accounting for the total number of available teeth\cite{anyaso2024testing}. Thus, an analysis of periodontal--caries association must potentially contend with informative structure both at the level of the mouth and within the disease-defined composition of that mouth.

Motivated by this gap, we develop two weights to be used in estimating marginal association between paired clustered outcomes in the presence of ISS. The first weight targets the association of all possible pairs within a population and is aptly named population pair weight (PPW). The other targets the association of pairs that are observed within the population, denoted observed pair weight (OPW). The proposed weights build on the WCR and weighted-estimation framework underlying CWGEEs, DWGEEs, and existing marginal association measures\cite{williamson2003marginal, huang2011informative, lorenz2018inferring}. The resulting construction allows scientifically meaningful categorizations of the paired outcomes to enter directly into the weighting scheme.

To evaluate the proposed weights, we develop a simulation framework that separately controls within-cluster unit-level association, latent cluster-level association, and outcome-dependent retention. This allows us to examine how the bidirectional subgroup structure defining ISS can arise under different sources of association and informative observation, and when pair-based weighting can improve estimation beyond conventional cluster weighting. We then apply the methods to tooth-level periodontal and caries data from the National Health and Nutrition Examination Survey (NHANES), where clinical attachment loss and pocket depth are paired with decayed- and filled-surface measures.

In Section~\ref{Preliminaries} we introduce the preliminary notation of GEEs and association measures. Following this, in Section~\ref{Methods} we discuss the methods used for weighting the GEEs, the simulation methodology, and ISS testing. Results are split into two sections, first we cover the results of the simulation in Section~\ref{Simulation}, and then in Section~\ref{Application} we show the results of the NHANES application. We close in Section~\ref{Discussion} with a discussion on the appropriate use of the weights under the simulated scenarios, limitations, and future work.

\section{Preliminaries}\label{Preliminaries}
\subsection{Weighted Generalized Estimating Equations}

Let $M$ be the number of clusters, indexed by $i$, and let $n_i$ be the number of observed units within cluster $i$, indexed by $j$. The total number of all observations can be defined as $N=\sum_{i=1}^{M}n_i$. Denote the bivariate response of observation $j$ within cluster $i$ as ($Y_{ij}$, $X_{ij}$). Typical notation represents each cluster by
\begin{equation*}
   \mathcal{V}_i=\{n_i,(X_{i1},Y_{i1}),\dots,(X_{in_i},Y_{in_i})\}. 
\end{equation*}
Assume that $\mathcal{V}_1,\dots,\mathcal{V}_M$ are independent and identically distributed. Under this assumption, the following weighted generalized estimating equation, denoted $\mathcal{U}$, may be used to estimate some scalar parameter of interest, denoted by $\theta$:
\begin{equation*}
    \mathcal{U}(\theta)
    =
    \sum_{i=1}^M \sum_{j=1}^{n_i}
    \omega_{ij} \mathcal{U}_{ij}(Y_{ij},X_{ij},\theta)
    =
    0,
\end{equation*}
where
\begin{equation*}
    E\{\mathcal{U}_{ij}(Y_{ij},X_{ij},\theta)\}=0,
\end{equation*}
and $\omega_{ij}$ is an arbitrary weight. The typical inverse cluster weighting is
\begin{equation}\label{eq:CW_weight}
\omega_{ij}^{CW}
=
\frac{1}{n_i}.
\end{equation}

The variance of the estimated scalar parameter, $\hat{\theta}$, is estimated in sandwich form by
\begin{equation*}
    \hat{\Sigma}
    =
    \frac{
    \hat{A}^{-1}
    \hat{B}
    \hat{A}^{-1}
    }{M},
\end{equation*}
where
\begin{align*}
    \hat{A}
    =
    \frac{1}{M}
    \sum_{i=1}^{M}
    \sum_{j=1}^{n_i}
    \omega_{ij}
    \left.
    \frac{
    \partial \mathcal{U}_{ij}(Y_{ij},X_{ij},\theta)
    }{
    \partial \theta
    }
    \right|_{\theta=\hat{\theta}}, \qquad
    \hat{B}
    =
    \frac{1}{M}
    \sum_{i=1}^{M}
    \left\{
    \sum_{j=1}^{n_i}
    \omega_{ij}
    \mathcal{U}_{ij}(Y_{ij},X_{ij},\hat{\theta})
    \right\}^{2}.
\end{align*}

One may also consider estimating a scalar function on $\theta$. If $g(\cdot)$ is a smooth scalar-valued function, then by the delta method, the variance of $g(\hat{\theta})$ is estimated by
\begin{equation*}
    \hat{\sigma}^2
    =
    \left[
    \hat{g}'(\hat{\theta})
    \right]^2
    \hat{\Sigma},
\end{equation*}
where
\begin{equation*}
    \hat{g}'(\hat{\theta})
    =
    \left.
    \frac{\partial g(\theta)}{\partial \theta}
    \right|_{\theta=\hat{\theta}}.
\end{equation*}

\subsection{Measures of Association}

Measures of association for clustered data have been developed for both paired and unpaired outcomes\cite{lorenz2011marginal,lorenz2018inferring}.
Here we focus on the paired setting, where both variables are observed on the same within-cluster unit.

For integers $k,l\ge 0$, define the marginal raw moment
$m_{kl}=E[X^kY^l]$. To estimate this moment, consider the scalar estimating
function
\begin{equation*}
    \mathcal{U}_{ij}(Y_{ij},X_{ij},\theta)
    =
    X_{ij}^kY_{ij}^l-\theta.
\end{equation*}
Since $E\{\mathcal{U}_{ij}(Y_{ij},X_{ij},m_{kl})\}=0$, solving the weighted estimating
equation
\begin{equation*}
    \sum_{i=1}^M \sum_{j=1}^{n_i}
    \omega_{ij}
    \left(X_{ij}^kY_{ij}^l-\hat{\theta}\right)
    =
    0
\end{equation*}
gives
\begin{equation}
    \hat{\theta}
    =
    \frac{
    \sum_{i=1}^M \sum_{j=1}^{n_i}
    \omega_{ij}X_{ij}^kY_{ij}^l
    }{
    \sum_{i=1}^M \sum_{j=1}^{n_i}
    \omega_{ij}
    }
    \equiv
    \hat{m}_{kl}.
\label{eq:moment_est_general}
\end{equation}

Let $\mathbf{m}=(m_{10},m_{01},m_{11},m_{20},m_{02})^{T}$. The usual product-moment correlation may be written as the smooth functional
\begin{equation}
g(\mathbf{m})
=
\frac{m_{11}-m_{10}m_{01}}
{\sqrt{m_{20}-m_{10}^2}\sqrt{m_{02}-m_{01}^2}}.
\label{eq:corr_fun}
\end{equation}
For continuous paired outcomes, $g(\mathbf{m})$ is the marginal Pearson correlation of the bivariate outcomes. By estimating $\hat{\mathbf{m}}=(\hat{m}_{10},\hat{m}_{01},\hat{m}_{11},\hat{m}_{20},\hat{m}_{02})^{T}$ from the weighted moment estimators in \eqref{eq:moment_est_general}, the estimator for $\rho_{\text{Pearson}}$ is then $g(\hat{\mathbf{m}})$\cite{lorenz2011marginal}.

Lorenz et al. show that this same formulation can be used to estimate the Spearman, chi, and other association measures. However, these measures were not utilized in this research. 

\section{Methods}\label{Methods}

\subsection{Categorization of outcomes}\label{categorization}

Let the paired continuous outcomes $X_{ij}$ and $Y_{ij}$ be mapped to categorical representations,

\begin{equation*}
X_{ij}\mapsto K_{ij}\in\{1,\dots,N_K\},
\qquad
Y_{ij}\mapsto L_{ij}\in\{1,\dots,N_L\}.
\end{equation*}

Here, $K_{ij}$ denotes the category assigned to $X_{ij}$, and similarly for $L_{ij}$ to $Y_{ij}$, for unit $j$ in cluster $i$. The quantities $N_K$ and $N_L$ denote the total numbers of possible categories used to represent the $X$ and $Y$ margins, respectively. The paired category of unit $j$ in cluster $i$ is

\begin{equation*}
P_{ij}=(K_{ij},L_{ij}),
\end{equation*}

which records the joint category membership of the two outcomes for a given unit. Let $\mathcal{P}$ denote the set of paired categories considered in the target population, and let

\begin{equation*}
N_P = |\mathcal{P}|
\end{equation*}

denote the total number of possible paired categories. If every combination of the $N_K$ categories of $X$ and the $N_L$ categories of $Y$ is allowable, then $N_P=N_KN_L$. Within cluster $i$, the set of paired categories actually observed as
\begin{equation*}
P_i=\{P_{ij}:j=1,\dots,n_i\},
\end{equation*}
and similarly let $N_{iP}=|P_i|$ denote the number of distinct paired categories represented among the observed units in cluster $i$. For unit $j$, 
\begin{equation*}
n_{i,P_{ij}}
=
\sum_{h=1}^{n_i}I(P_{ih}=P_{ij}),
\end{equation*}

where $n_{i,P_{ij}}$ is the number of units in cluster $i$ belonging to the same paired category as unit $j$. For notational simplicity, we write the latter as $n_{iP}$ when the unit under consideration is clear. The categorization is used only to construct the weights. Association is estimated from the original paired outcomes $(X_{ij},Y_{ij})$ using the weighted estimating equations defined in Section~\ref{Preliminaries}.

\subsection{Weights}\label{section:weights}

The proposed weights are motivated by within-cluster resampling and the corresponding weighted estimating equation arguments used for informative cluster and subgroup sizes \cite{williamson2003marginal,huang2011informative}. Here, we define the weighting schemes directly in terms of the paired outcome categories used in the resampling construction. The corresponding resampling arguments, derivations of the weights, and asymptotic representation are provided in the Supplementary Materials.

\subsubsection{Population pair weights (PPW)}

Population pair weighting begins with the target population set $\mathcal{P}$. For each cluster $i$, consider the following resampling scheme:

\begin{enumerate}
    \item Select one paired category uniformly from the $N_P$ categories in $\mathcal{P}$.
    \item If that paired category is observed in cluster $i$, select one unit uniformly from the units belonging to that category.
    \item If the paired category is not observed in cluster $i$, record a null draw.
\end{enumerate}

For an observed unit $j$, the probability of selection is therefore

\begin{equation*}
\pi_{ij}^{PPW}
=
\frac{1}{N_P}
\frac{1}{n_{iP}}.
\end{equation*}

The probability of a null draw from cluster $i$ is

\begin{equation*}
1-\frac{N_{iP}}{N_P}.
\end{equation*}

Because $N_P$ is fixed across clusters, the factor $1/N_P$ is common to the entire estimating equation and does not affect its solution. The resulting population pair weight is therefore

\begin{equation}\label{eq:PPW_weight}
\omega_{ij}^{PPW}
=
\frac{1}{n_{iP}}.
\end{equation}

Under PPW, every paired category observed within a cluster contributes equally regardless of how many units belong to that category. Categories in the target population that are absent from a particular cluster are represented by null draws rather than by imputing or modeling unobserved outcomes. Consequently, the total observed contribution of a cluster depends on the number of paired categories represented within that cluster.

\subsubsection{Observed pair weights (OPW)}

Observed pair weighting instead restricts the resampling distribution to the paired categories actually observed within a cluster. For cluster $i$:

\begin{enumerate}
    \item Select one paired category uniformly from the $N_{iP}$ categories in $P_i$.
    \item Select one unit uniformly from the units belonging to the selected category.
\end{enumerate}

The probability of selecting observed unit $j$ is

\begin{equation*}
\pi_{ij}^{OPW}
=
\frac{1}{N_{iP}}
\frac{1}{n_{iP}},
\end{equation*}

giving the weight

\begin{equation}\label{eq:OPW_weight}
\omega_{ij}^{OPW}
=
\frac{1}{N_{iP}n_{iP}}.
\end{equation}

OPW therefore gives equal total weight to each observed paired category and, because

\begin{equation*}
\sum_{j=1}^{n_i}\omega_{ij}^{OPW}=1,
\end{equation*}

also gives equal total weight to each cluster. PPW and OPW coincide up to a common multiplicative constant when every cluster contains the same number of paired categories; they differ when the representation of paired categories varies across clusters.

\subsection{Simulation}

To evaluate the finite-sample performance of the proposed weighted estimators, we generated clustered paired outcomes under a latent Gaussian model with informative observation. For each cluster, we first generated latent cluster-level variables
\begin{equation*}
(U_i,V_i)^\top \sim N_2\!\left(
\begin{pmatrix}
\mu_u\\
\mu_v
\end{pmatrix},
\begin{pmatrix}
\sigma_u^2 & \rho_{uv}\sigma_u\sigma_v\\
\rho_{uv}\sigma_u\sigma_v & \sigma_v^2
\end{pmatrix}
\right),
\end{equation*}
where $U_i$ and $V_i$ represent cluster-level latent effects governing the two paired outcome margins.

Conditional on $(U_i,V_i)$, we then generated $n_{\max}$ potential paired outcomes
\begin{equation*}\label{sim:model}
(X_{ij},Y_{ij})^\top \sim
N_2\!\left(
\begin{pmatrix}
\alpha_x+\beta_x U_i\\
\alpha_y+\beta_y V_i
\end{pmatrix},
\begin{pmatrix}
\sigma_x^2 & \rho_{xy}\sigma_x\sigma_y\\
\rho_{xy}\sigma_x\sigma_y & \sigma_y^2
\end{pmatrix}
\right),
\qquad j=1,\dots,n_{\max}.
\end{equation*}
Hence, $\rho_{uv}$ induces dependence among units within-cluster, while $\rho_{xy}$ governs the unit association. These two sources determine the full-data marginal correlation, denoted by $\rho_0$. The corresponding derivation and closed-form expression for $\rho_0$ are provided in the Supplementary Materials.

\subsubsection{Categorization of simulated unit-level variables}

Using the categorical notation introduced in Subsection~\ref{categorization}, we mapped each simulated continuous margin to its corresponding ordinal category. Specifically, $X_{ij}$ was mapped to $K_{ij}\in\{1,\dots,N_K\}$ and $Y_{ij}$ was mapped to $L_{ij}\in\{1,\dots,N_L\}$ by discretizing standardized versions of the two margins.

We first standardized each continuous margin using its unconditional mean and variance:
\begin{equation*}
X_{ij}^{*}
=
\frac{X_{ij}-(\alpha_x+\beta_x\mu_u)}
{\sqrt{\beta_x^2\sigma_u^2+\sigma_x^2}},
\qquad
Y_{ij}^{*}
=
\frac{Y_{ij}-(\alpha_y+\beta_y\mu_v)}
{\sqrt{\beta_y^2\sigma_v^2+\sigma_y^2}}.
\end{equation*}
We then applied equal-probability standard normal quantile cuts separately to the two margins. For each margin, define
\begin{equation*}
c_{(\cdot),h}
=
\Phi^{-1}\!\left(\frac{h}{N_{(\cdot)}}\right),
\qquad h=1,\dots,N_{(\cdot)}-1,
\end{equation*}
with $c_{(\cdot),0}=-\infty$ and $c_{(\cdot),N_{(\cdot)}}=+\infty$. The ordinal categories were then defined by
\begin{equation*}
K_{ij}=h
\quad \text{if} \quad
c_{K,h-1}\le X_{ij}^{*}<c_{K,h},
\qquad h=1,\dots,N_K,
\end{equation*}
and
\begin{equation*}
L_{ij}=h
\quad \text{if} \quad
c_{L,h-1}\le Y_{ij}^{*}<c_{L,h},
\qquad h=1,\dots,N_L.
\end{equation*}
Since standardization used the model-implied unconditional mean and variance, the same standard-normal quantile cutpoints were used across replicates within each simulation setting.

\subsubsection{Observation retention}

To induce informative cluster size, each potential unit was assigned an observation indicator $\mathcal{R}_{ij}\in\{0,1\}$ according to a logistic retention model depending on the continuous paired outcomes:
\begin{equation*}
\Pr(\mathcal{R}_{ij}=1\mid X_{ij},Y_{ij})
=
\min\left\{
\operatorname{logit}^{-1}\!\left(\eta_0+\eta_xX_{ij}\right),
\operatorname{logit}^{-1}\!\left(\eta_0+\eta_yY_{ij}\right)
\right\}.
\end{equation*}
Hence,
\begin{equation*}
\mathcal{R}_{ij}\sim \operatorname{Bernoulli}\!\left(
\Pr(\mathcal{R}_{ij}=1\mid X_{ij},Y_{ij})
\right).
\end{equation*}
Larger positive values of $\eta_x$ or $\eta_y$ produced stronger dependence of retention on the corresponding outcome margin and, consequently, greater outcome-dependent variation in observed cluster size. For each simulation replicate, we independently generated $M$ clusters according to the above mechanism. The observed cluster consisted of the realized units satisfying $\mathcal{R}_{ij}=1$, with realized cluster size
\begin{equation*}
n_i^{\text{obs}}
=
\sum_{j=1}^{n_{\max}}\mathcal{R}_{ij}.
\end{equation*}
Thus, inference was based only on retained units from clusters satisfying $n_i^{\text{obs}}\geq n_{\min}$, where $n_{\min}$ is the minimum number of observations required to calculate association measures.

\subsubsection{Performance summary}
%\textcolor{blue}{\Large Just going to choose one of these two, most likely the True coverage.}

Estimator performance was summarized by the Monte Carlo average of the estimated correlation, the 
empirical coverage of nominal 95\% confidence interval, and the absolute bias from the full-data marginal correlation. Let $\hat{\rho}^{(q)}$ denote the estimated correlation in replicate $q=1,\dots,Q$, and let $\hat{SE}^{(q)}$ denote the corresponding sandwich and delta-method standard error. We computed the average of the estimated correlation as
\begin{equation*}
\overline{\rho}
=
\frac{1}{Q}\sum_{q=1}^Q \hat{\rho}^{(q)}.
\end{equation*}

Let $\rho_0$ denote the full-data marginal correlation under the simulation setting. Absolute bias was calculated as

\begin{equation*}
\operatorname{Absolute\ Bias}
=
\left|\overline{\rho}-\rho_0\right|.
\end{equation*}

Let $\rho_0$ denote the full-data marginal correlation under the simulation setting. The empirical coverage of $\rho_0$ was computed as
\begin{equation*}
\mathcal{C}_{0.95}^{\text{true}}
=
\frac{1}{Q}
\sum_{q=1}^Q
I\left[
\hat{\rho}^{(q)}
-
z_{0.975}\hat{SE}^{(q)}
\le
\rho_0
\le
\hat{\rho}^{(q)}
+
z_{0.975}\hat{SE}^{(q)}
\right],
\end{equation*}
where \(z_{0.975}\) denotes the 97.5th percentile of the standard normal distribution.

\subsection{Marginal Association Testing Under Informative Subgroup Size}
\label{section:ISS_test}

To assess the presence of ISS, we use the testing procedure of Anyaso-Samuel and Datta (2024), implemented by the function \verb|ASDpv| in the \verb|crspack| package \cite{anyaso2024testing}. The procedure was developed for clustered data in which a continuous unit-level variable induces latent within-cluster subgroups whose sizes may themselves be informative. For a response $Y_{ij}$ and subgroup-inducing variable $Z_{ij}$, latent binary subgroups are formed across thresholds $z$ of $Z_{ij}$, and the corresponding marginal distributions of $Y_{ij}$ are compared using subgroup-size-adjusted rank statistics \cite{anyaso2024testing,dutta2016rank}. The resulting statistic tests whether the marginal distribution of $Y_{ij}$ varies across subgroups induced by $Z_{ij}$ while accounting for the induced subgroup sizes. In our terminology, rejection provides evidence of ISS in the direction from $Z_{ij}$ to $Y_{ij}$.

For the paired outcomes considered here, we apply the procedure in both directions. First, $Y_{ij}$ is treated as the response and $X_{ij}$ as the subgroup-inducing variable. We then reverse their roles, treating $X_{ij}$ as the response and $Y_{ij}$ as the subgroup-inducing variable. We refer to ISS for paired outcomes as the bidirectional case in which both tests reject, indicating that the subgroups induced by each outcome are informative for the marginal distribution of the other.

The original procedure obtains inference through within-cluster permutation of the subgroup-inducing variable \cite{anyaso2024testing}. Because the permutation calculation becomes computationally demanding for data containing large numbers of clusters, we partition the clusters into small, disjoint subsets and perform the permutation procedure separately within each subset. The resulting subset-specific p-values are then combined using Stouffer's inverse-normal method \cite{stouffer1949american}. The details of this computational implementation are provided in the Supplementary Materials.
\FloatBarrier
\section{Simulation Results}\label{Simulation}

We evaluated the proposed estimators across simulation settings designed to distinguish the source of association from the source of informative observation. The parameters $\rho_{uv}$ and $\rho_{xy}$ control association arising through correlated cluster-level latent effects and paired unit-level outcomes, respectively, while $\eta_x$ and $\eta_y$ determine whether retention depends on the $X$ or $Y$ outcome margin. The resulting simulation grid is shown in Table~\ref{tab:sim_grid}.

\begin{table}[htbp]
\centering
\small
\begin{tabular}{lll}
\toprule
Parameter & Values & Description \\
\midrule
$\rho_{uv}$ & $\{0,\ 0.5\}$ & Correlation between cluster-level latent variables \\
$\rho_{xy}$ & $\{0,\ 0.5\}$ & Within-unit correlation between paired outcomes \\
$\eta_x$ & $\{0,\ 4\}$ & Strength of retention dependence on the $X$ margin \\
$\eta_y$ & $\{0,\ 4\}$ & Strength of retention dependence on the $Y$ margin \\
\bottomrule
\end{tabular}
\caption{Simulation parameter grid. The parameters shown were varied factorially across simulation settings.}
\label{tab:sim_grid}
\end{table}

The varied parameters form a full factorial grid of $16$ simulation settings. Across all settings, we fixed
$M=20$,
$(\mu_u,\mu_v)=(0,0)$,
$(\sigma_u,\sigma_v)=(1,1)$,
$(\alpha_x,\alpha_y)=(0,0)$,
$(\beta_x,\beta_y)=(1,1)$,
$(\sigma_x,\sigma_y)=(0.5,0.5)$,
$(N_K,N_L)=(5,5)$,
$\eta_0=3$,
$n_{\max}=100$, and
$n_{\min}=2$.
For each setting, we generated $Q=10{,}000$ independent Monte Carlo replicates. Following observation retention, each estimator was applied to the retained observations, with clusters contributing to the association analysis only when at least $n_{\min}=2$ units remained.

\subsection{Informative Subgroup Testing Results}

The ISS testing results in Table~\ref{tab:sim_stouffer} were computed from the full collection of Monte Carlo simulated datasets. Within each individual test, we used subsets of 10 clusters and the resulting $p$-values were then combined using Stouffer's method. Because the clusters are independently generated across Monte Carlo replicates, the Stouffer combination is being applied over independent cluster-level contributions, so no additional dependence correction is required for that aggregation.

\begin{table}[!ht]
\small
\centering
\setlength{\tabcolsep}{6pt}
\renewcommand{\arraystretch}{0.95}
\begin{tabular}{llll c l c ll}
\toprule
\multicolumn{4}{l}{Parameters} && True Corr. && \multicolumn{2}{c}{Direction}\\
\cmidrule{1-4} \cmidrule{6-6} \cmidrule{8-9}
$\rho_{xy}$ & $\rho_{uv}$ & $\eta_x$ & $\eta_y$
&& $\rho_0$ && $X$ & $Y$\\
\midrule
0.5 & 0.5 & 0 & 0 && $0.5$ && $<10^{-16}$ & $<10^{-16}$\\
0.5 & 0.5 & 0 & 4 && $0.5$ && $<10^{-16}$ & $<10^{-16}$\\
0.5 & 0.5 & 4 & 0 && $0.5$ && $<10^{-16}$ & $<10^{-16}$\\
0.5 & 0.5 & 4 & 4 && $0.5$ && $<10^{-16}$ & $<10^{-16}$\\
\addlinespace[4pt]

0.5 & 0   & 0 & 0 && $0.1$ && $<10^{-16}$ & $<10^{-16}$\\
0.5 & 0   & 0 & 4 && $0.1$ && $<10^{-16}$ & $<10^{-16}$\\
0.5 & 0   & 4 & 0 && $0.1$ && $<10^{-16}$ & $<10^{-16}$\\
0.5 & 0   & 4 & 4 && $0.1$ && $<10^{-16}$ & $<10^{-16}$\\
\addlinespace[4pt]

0   & 0.5 & 0 & 0 && $0.4$ && $0.3527$ & $0.3765$\\
0   & 0.5 & 0 & 4 && $0.4$ && $0.0096$ & $0.1226$\\
0   & 0.5 & 4 & 0 && $0.4$ && $0.6446$ & $0.0429$\\
0   & 0.5 & 4 & 4 && $0.4$ && $0.0006$ & $0.0002$\\
\addlinespace[4pt]

0   & 0   & 0 & 0 && $0$ && $0.8390$ & $0.9304$\\
0   & 0   & 0 & 4 && $0$ && $0.0956$ & $0.1202$\\
0   & 0   & 4 & 0 && $0$ && $0.3460$ & $0.1318$\\
0   & 0   & 4 & 4 && $0$ && $2.0764\times10^{-7}$ & $1.3170\times10^{-11}$\\
\bottomrule
\end{tabular}
\captionsetup{width=0.95\linewidth}
\caption{Informative subgroup size tests for the simulation study. Entries are Stouffer combined $p$-values from tests using independent subsets of size 10 and all possible permutations. ``$X$'' denotes tests in which $X_{ij}$ is the subgroup-defining variable, whereas ``$Y$'' denotes tests in which $Y_{ij}$ is the subgroup-defining variable.}
\label{tab:sim_stouffer}
\end{table}

When the outcome correlation is present at the unit-level, that is when $\rho_{xy}=0.5$, there is evidence of informative subgroup size in both directions, regardless of the value of the latent cluster-level correlation, $\rho_{uv}$. The strongest driver of the observed subgroup informativeness is seen to be the within-unit association, even though the true value of association is higher when the latent correlation is present while the unit-level correlation is not. 

When both unit-level retention variables are present, $\eta_x=4, \eta_y=4$, the ISS test rejects strongly in both directions; a consequence of the observation mechanism. When both retention parameters are large, lower values of both $X_{ij}$ and $Y_{ij}$ are associated with reduced observation, which changes the observed within-cluster composition regardless of the true correlation. This shows that ISS may arise from outcome-dependent observation acting simultaneously on both margins. Additionally, when only one retention parameter is present, the evidence is direction-specific. This is consistent with the interpretation of the ISS test as a directional diagnostic. 

These results suggest that the ISS test is highly sensitive to direct within-unit association, less sensitive to latent cluster-level correlation by itself, and best detects subgroup-size informativeness induced purely through the observation process.

\subsection{Association Measure Results}

Table~\ref{tab:sim_assoc_M20} summarizes the performance of the weighted Pearson correlation estimators. For each weighting scheme, we report the Monte Carlo mean estimate and empirical coverage of the nominal 95\% confidence interval for the complete-data target.

\begin{table}[!ht]
    \centering
    \small
    \setlength{\tabcolsep}{3pt}
    \begin{tabular*}{0.95\linewidth}{@{\extracolsep{\fill}}llll c r c rrr}
       \toprule
       \multicolumn{4}{l}{Params.} && \multicolumn{1}{l}{Target} && \multicolumn{3}{l}{Association by Weight: $\hat{\rho}\,(\mathcal{C}_{0.95}^{true})$}\\
       \cmidrule{1-4} \cmidrule{6-6} \cmidrule{8-10}
       $\rho_{xy}$ & $\rho_{uv}$ & $\eta_x$ & $\eta_y$
       && True && CW & PPW & OPW \\
       \cmidrule{1-4} \cmidrule{6-6} \cmidrule{8-10}
       0.5 & 0.5 & 0 & 0 && $0.5$ && $0.49\,(0.87)$ & $0.32\,(0.33)$ & $0.37\,(0.67)$ \\
       0.5 & 0.5 & 0 & 4 && $0.5$ && $0.41\,(0.87)$ & $0.30\,(0.31)$ & $0.34\,(0.64)$ \\
       0.5 & 0.5 & 4 & 0 && $0.5$ && $0.42\,(0.87)$ & $0.30\,(0.32)$ & $0.34\,(0.65)$ \\
       0.5 & 0.5 & 4 & 4 && $0.5$ && $0.41\,(0.86)$ & $0.31\,(0.42)$ & $0.34\,(0.69)$ \\
       \addlinespace
       0.5 & 0 & 0 & 0 && $0.1$ && $0.11\,(0.87)$ & $0.14\,(0.89)$ & $0.10\,(0.88)$ \\
       0.5 & 0 & 0 & 4 && $0.1$ && $0.05\,(0.87)$ & $0.12\,(0.91)$ & $0.06\,(0.88)$ \\
       0.5 & 0 & 4 & 0 && $0.1$ && $0.05\,(0.87)$ & $0.12\,(0.91)$ & $0.06\,(0.88)$ \\
       0.5 & 0 & 4 & 4 && $0.1$ && $0.03\,(0.85)$ & $0.12\,(0.91)$ & $0.04\,(0.86)$ \\
       \addlinespace
       0 & 0.5 & 0 & 0 && $0.4$ && $0.38\,(0.88)$ & $0.15\,(0.07)$ & $0.25\,(0.61)$ \\
       0 & 0.5 & 0 & 4 && $0.4$ && $0.36\,(0.88)$ & $0.16\,(0.14)$ & $0.26\,(0.68)$ \\
       0 & 0.5 & 4 & 0 && $0.4$ && $0.36\,(0.88)$ & $0.16\,(0.14)$ & $0.26\,(0.68)$ \\
       0 & 0.5 & 4 & 4 && $0.4$ && $0.38\,(0.88)$ & $0.18\,(0.27)$ & $0.28\,(0.75)$ \\
       \addlinespace
       0 & 0 & 0 & 0 && $0$ && $0\,(0.88)$ & $0\,(0.93)$ & $0\,(0.89)$ \\
       0 & 0 & 0 & 4 && $0$ && $0\,(0.88)$ & $0\,(0.92)$ & $0\,(0.89)$ \\
       0 & 0 & 4 & 0 && $0$ && $0\,(0.87)$ & $0\,(0.92)$ & $0\,(0.88)$ \\
       0 & 0 & 4 & 4 && $0$ && $0.04\,(0.87)$ & $0.02\,(0.91)$ & $0.02\,(0.88)$ \\
       \bottomrule
    \end{tabular*}%
    \captionsetup{width=0.95\linewidth}
    \caption{Simulation results for Pearson correlation. Entries under each weighting scheme are estimated associations, with empirical coverage probability for $\rho_0$ reported in parentheses. All simulated severity levels were taken as categorical values for weighting.}
    \label{tab:sim_assoc_M20}
\end{table}

When association is present only at the unit level ($\rho_{xy}=0.5,\rho_{uv}=0$), PPW performs best overall. Its estimates remain closest to the target as retention is introduced, while CW and OPW become more attenuated. This agrees with the ISS testing results in Table~\ref{tab:sim_stouffer}, where these settings produced strong evidence of informative subgroup size in both directions.

When association is present only at the cluster level ($\rho_{xy}=0,\rho_{uv}=0.5$), CW remains closest to the target of $0.4$, while PPW and OPW substantially underestimates the association. These settings also produced much weaker evidence of ISS unless informativeness was introduced through the retention mechanism.

When both unit-level and cluster-level association are present ($\rho_{xy}=\rho_{uv}=0.5$), the ISS tests again strongly reject in both directions. However, CW remains closest to the target of $0.5$, while PPW and OPW underestimate the association.

When neither source of association is present, all three estimators remain close to zero. However, in the same setting when both retention parameters are large ($\eta_x=\eta_y=4$), a small positive association appears, once again demonstrating that the observation mechanism alone can induce apparent association.

The results are explained by the different weighting targets. CW equalizes the total contribution of each cluster, and PPW additionally equalizes the contribution of the paired response categories within each cluster. This is beneficial when the association is primarily generated through the within-unit relationship between $X$ and $Y$, but it can attenuate association generated through latent between-cluster variation. OPW produces an intermediate degree of reweighting because it balances only the paired categories observed within each cluster.

\begin{figure}[!ht]
    \centering
    \includegraphics[width=\linewidth]{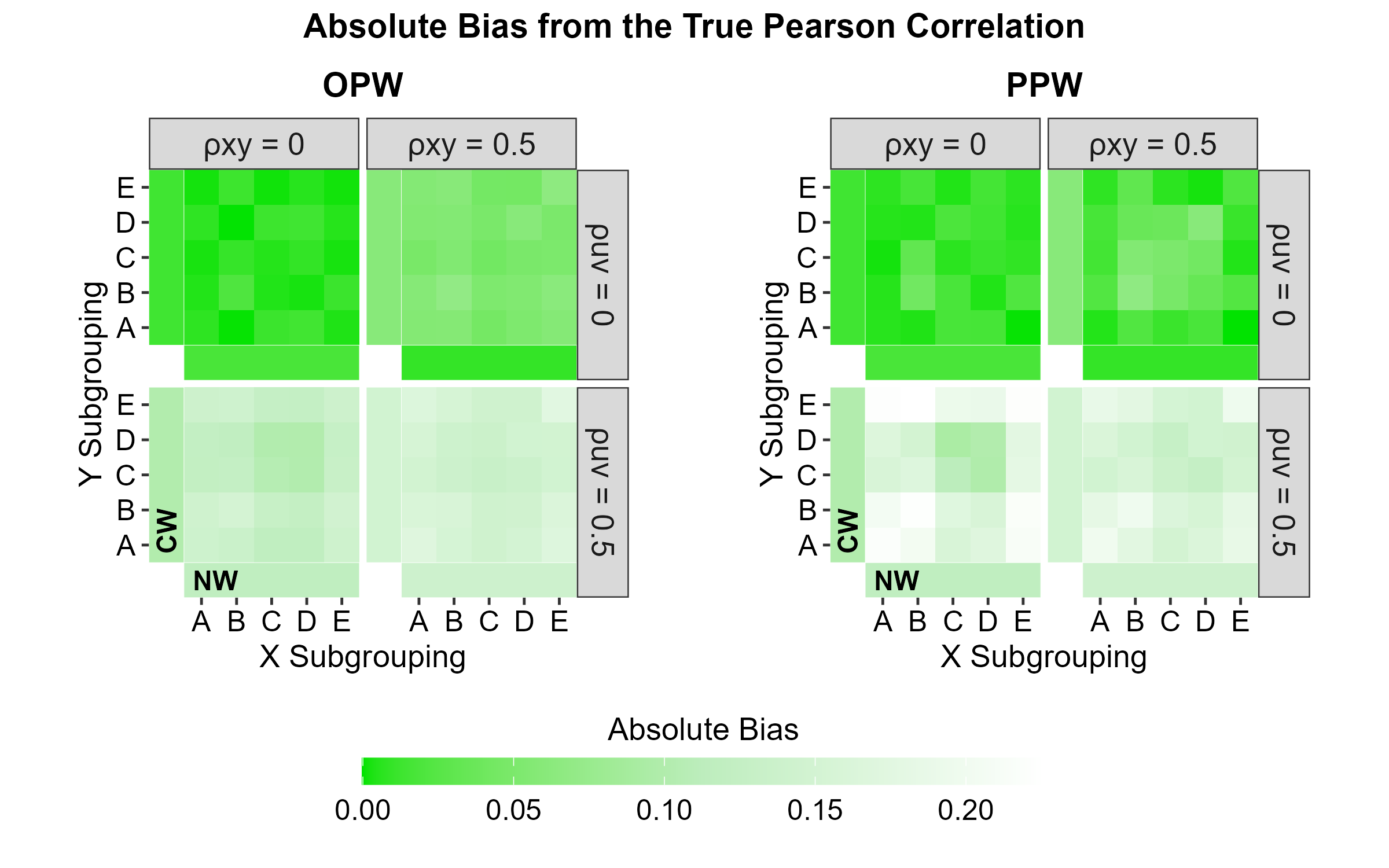}

    \caption{Absolute bias of the Pearson correlation estimators under alternative categorizations of the paired responses. The left and right panels correspond to outcome probability weighting (OPW) and pairwise probability weighting (PPW), respectively. Within each panel, columns correspond to values of $\rho_{xy}$ and rows correspond to values of $\rho_{uv}$, with $\eta_x=\eta_y=4$. Within each subplot, rows and columns correspond to alternative dichotomizations of the categories of $Y$ and $X$, respectively, where the vertical line indicates a split between low and high. Labels A--D denote the four dichotomizations, while E denotes the full Severity categorization equivalent to the values in Table~\ref{tab:sim_assoc_M20}. The bottom and left bars represent the no-weighting (NW) and cluster-weighting (CW) estimators, respectively. Smaller values shaded green indicate lower absolute bias relative to the complete-data target correlation.}
    \label{fig:heatmap}
\end{figure}

% Heatmap Results
We are interested in how the dichotomization of the paired response categories affects the resulting correlation estimates. Figure~\ref{fig:heatmap} illustrates this setting in which we also retain the retention mechanism ($\eta_x=\eta_y=4$). The exact values within Figure~\ref{fig:heatmap} are available in downloadable materials within the Supplementary Materials. For both $X$ and $Y$, five subgroup definitions were considered: A separates category 1 from 2--5, B separates 1--2 from 3--5, C separates 1--3 from 4--5, D separates 1--4 from 5, and E retains the full five-level severity categorization. Thus, A--D represent successive dichotomizations obtained by moving the cutpoint across the ordered response categories, while E retains the original categorical structure. Because NW and CW do not depend on the subgroup definition, their absolute bias is constant across these categorizations, whereas PPW and OPW vary as the subgroup definitions for the paired responses change.

When $\rho_{uv}=0$, OPW and PPW generally have similar absolute bias across the categorizations when compared to NW and CW. PPW performs particularly well for the more extreme splits, with some combinations involving subgroup A producing little or no absolute bias. This is consistent with the earlier simulation results, in which pair-based weighting performs well when the association is primarily generated at the unit level.

When $\rho_{xy}=0$ and $\rho_{uv}=0.5$, CW generally has lower absolute bias than NW and many of the PPW and OPW categorizations. However, the performance of the pair-based weights depends strongly on the chosen dichotomizations. Both OPW and PPW improve substantially for combinations involving C and D, corresponding to splits that isolates the most severe categories. In contrast, categorizations involving A and B can produce substantially greater bias than either CW or NW. This is more pronounced for PPW than OPW, however OPW offers more consistency. 

When both unit- and cluster-level associations are present, with $\rho_{xy}=0.5$ and $\rho_{uv}=0.5$, the same general pattern persists. C and D tend to produce lower absolute bias, whereas A and B tend to perform less favorably. Once again, this dependence on categorization is more pronounced for PPW, for which the differences between well- and poorly-performing dichotomizations are larger than those observed for OPW. 

The paired weights therefore introduce a choice that is absent from CW: the subgroup definition itself affects the estimated association. When the categorization aligns well with the subgroup imbalance, PPW or OPW can substantially reduce bias; in this case, for dichotomizations that isolates the most severe categories which are subject to a retention mechanism.

\FloatBarrier
\section{Application to NHANES}\label{Application}

We applied the proposed weighted association estimators to tooth-level data from the 2011--2014 NHANES, introduced in Section~\ref{intro}. NHANES collects data from a nationally representative sample of non-institutionalized U.S. residents through household interviews and standardized physical examinations conducted in mobile examination centers. The oral health component contains repeated tooth-level measurements within individuals, providing a natural setting for evaluating marginal association in clustered paired outcomes where both cluster and subgroup sizes may be informative.

Individuals aged 30 years or older who completed the oral health examination were considered. Third molars were excluded, and analyses were restricted to individuals with more than 9 observed teeth. Periodontal and caries measurements were summarized at the tooth level so that each observed tooth contributed a paired periodontal--caries outcome.

The periodontal outcomes were clinical attachment loss (CAL) and pocket depth (PD), each summarized by the maximum valid measurement across periodontal sites within a tooth. Caries experience was represented using six tooth-level surface scores: decayed surfaces (DS), filled surfaces (FS), decayed or filled surfaces (DFS), and their corresponding interproximal measures DSI, FSI, and DFSI. Each score represented the proportion of eligible tooth surfaces exhibiting the corresponding condition, based on the surface-level caries classifications defined in the NHANES oral health examination protocol\cite{nhanes2011_2014}.

We considered 12 periodontal--caries pairings, with CAL and PD each paired with DS, DSI, FS, FSI, DFS, and DFSI. Because both members of each pair were treated as continuous tooth-level outcomes, the analyses focused on marginal Pearson correlation. The weighting schemes described in Section~\ref{section:weights} were then used to evaluate how adjustment for informative cluster and subgroup sizes affected the estimated associations.

\subsection{NHANES Survey Weighting}

To account for the complex NHANES sampling design, person-level MEC examination weights were incorporated into all association analyses. Because the 2011--2012 and 2013--2014 cycles were combined, the four-year examination weight was defined as
\[
W_i^{\mathrm{NHANES}}
=
\frac{WTMEC2YR_i}{2}.
\]
The supplied NHANES weights are already post-stratified to U.S. population totals and therefore required no additional adjustment\cite{chen2018national}. The NHANES sampling weight was combined multiplicatively with the informative cluster- or subgroup-size weight. Thus, for tooth \(j\) from participant \(i\), the weight used in the estimating equation was
\[
w_{ij}
=
W_i^{\mathrm{NHANES}}w_{ij}^{*},
\]
where \(w_{ij}^{*}\) denotes the corresponding NW, CW, PPW, or OPW contribution. This allows the association estimates to account simultaneously for the NHANES sampling design and informative tooth-level cluster structure.

As a sensitivity analysis, we also repeated the analyses without the NHANES sampling weights. The resulting estimates were similar, with the same direction of association and the same conclusions regarding significance.

\FloatBarrier
\begin{table}[htbp]
\centering
\begin{tabular}{l cc cc}
\toprule
& \multicolumn{2}{c}{CAL} & \multicolumn{2}{c}{PD}\\
\cmidrule(lr){2-3} \cmidrule(lr){4-5}
Caries outcome
& Perio & Caries
& Perio & Caries\\
\midrule
DS   & $1$         & $0.9153$      & $1$         & $0.5945$\\
DSI  & $1$         & $0.5492$      & $1$         & $0.3148$\\
\addlinespace[4pt]
FS   & $<10^{-16}$ & $<10^{-16}$  & $<10^{-16}$ & $<10^{-16}$\\
FSI  & $<10^{-16}$ & $<10^{-16}$  & $<10^{-16}$ & $<10^{-16}$\\
\addlinespace[4pt]
DFS  & $<10^{-16}$ & $<10^{-16}$  & $<10^{-16}$ & $<10^{-16}$\\
DFSI & $<10^{-16}$ & $<10^{-16}$  & $0.0087$     & $<10^{-16}$\\
\bottomrule
\end{tabular}
\captionsetup{width=0.95\linewidth}
\caption{Informative subgroup size tests for NHANES periodontal and caries outcomes. Entries are Stouffer combined $p$-values from tests using independent subsets of size 10 and all possible permutations. ``Perio'' denotes $Z$ equal to the periodontal outcome, whereas ``Caries'' denotes $Z$ equal to the corresponding zero-versus-nonzero caries measure.}
\label{tab:iss_tests}
\end{table}

Before comparing the resulting association estimates, we assessed whether the within-mouth subgroup sizes induced by one member of a paired outcome were informative for the marginal distribution of the other member. This was done using the ISS test described in Section~\ref{section:ISS_test}, with independent subsets of size 10. When the periodontal outcome defined the subgroups, CAL was dichotomized at 3, 4, and 5 mm and PD at 4, 5, and 6 mm, and When the caries outcome defined the subgroups\cite{aldosari2020depression}. The corresponding tooth-level measure was dichotomized as zero versus nonzero. Results are reported in Table~\ref{tab:iss_tests}.

The ISS test results showed clear differences across the caries outcomes. There was little evidence of informative subgroup size for either DS or DSI in either direction. In contrast, all pairings involving FS, FSI, or DFS showed strong evidence of informative subgroup size in both directions, with $p<10^{-16}$ for both CAL and PD. DFSI showed similarly strong evidence for CAL in both directions and when DFSI defined the PD subgroups, while the test in which PD defined the subgroup yielded $p=0.0087$. Consistent with the simulation results, these findings indicate that the subgroup-size structure differs substantially across the caries outcomes and motivates the use of subgroup-adjusted weighting in the association analyses.

\FloatBarrier
\begin{figure}[!ht]
    \centering
    \includegraphics[width=0.9\linewidth]{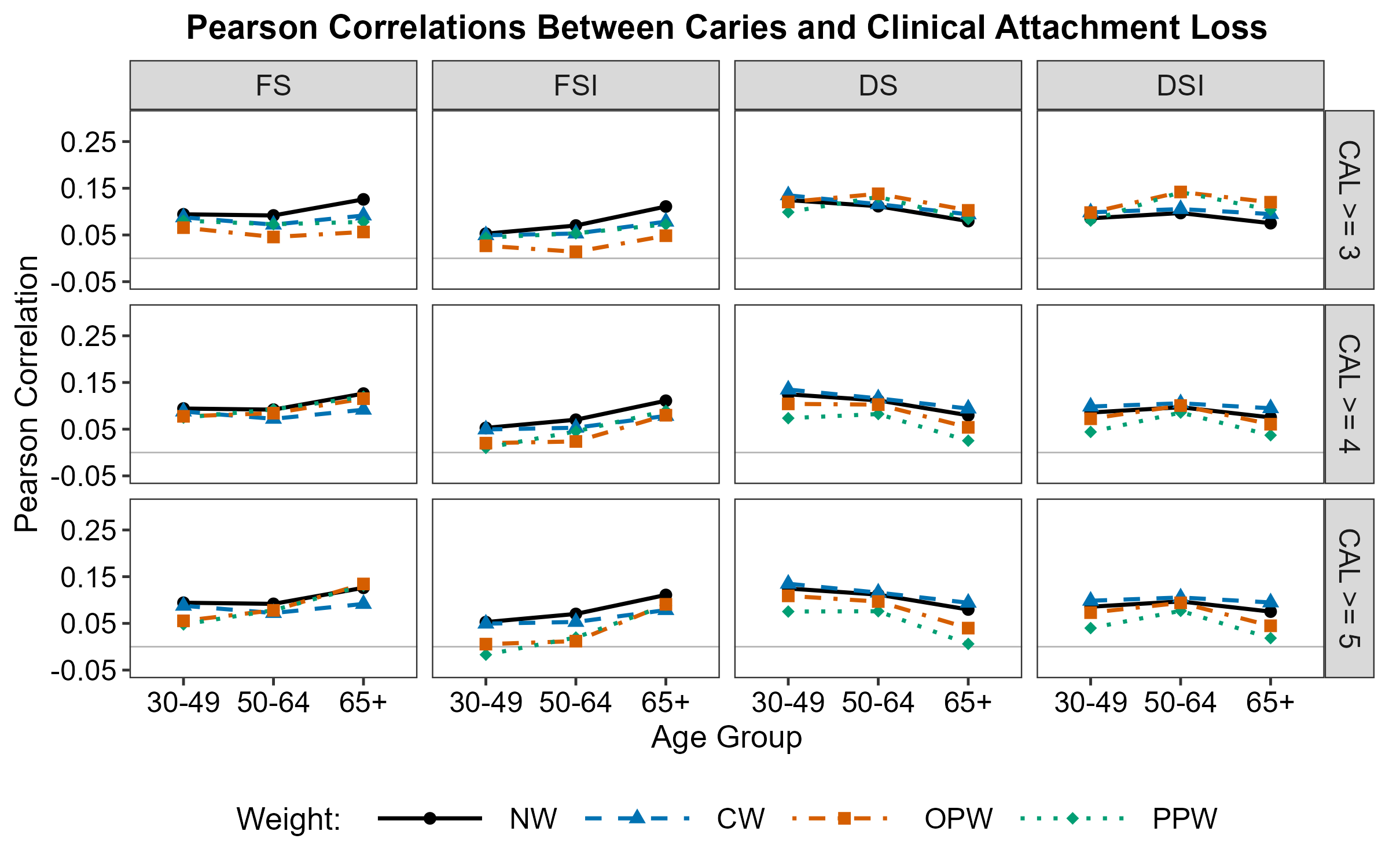}

    \captionsetup{width=0.95\linewidth}
    \caption{Pearson correlations between clinical attachment loss (CAL) and DS, DSI, FS, and FSI in NHANES across age groups and weighting methods. Rows correspond to increasing CAL thresholds.}
    \label{fig:calplots}
\end{figure}

\begin{figure}[!ht]
    \centering
    \includegraphics[width=0.9\linewidth]{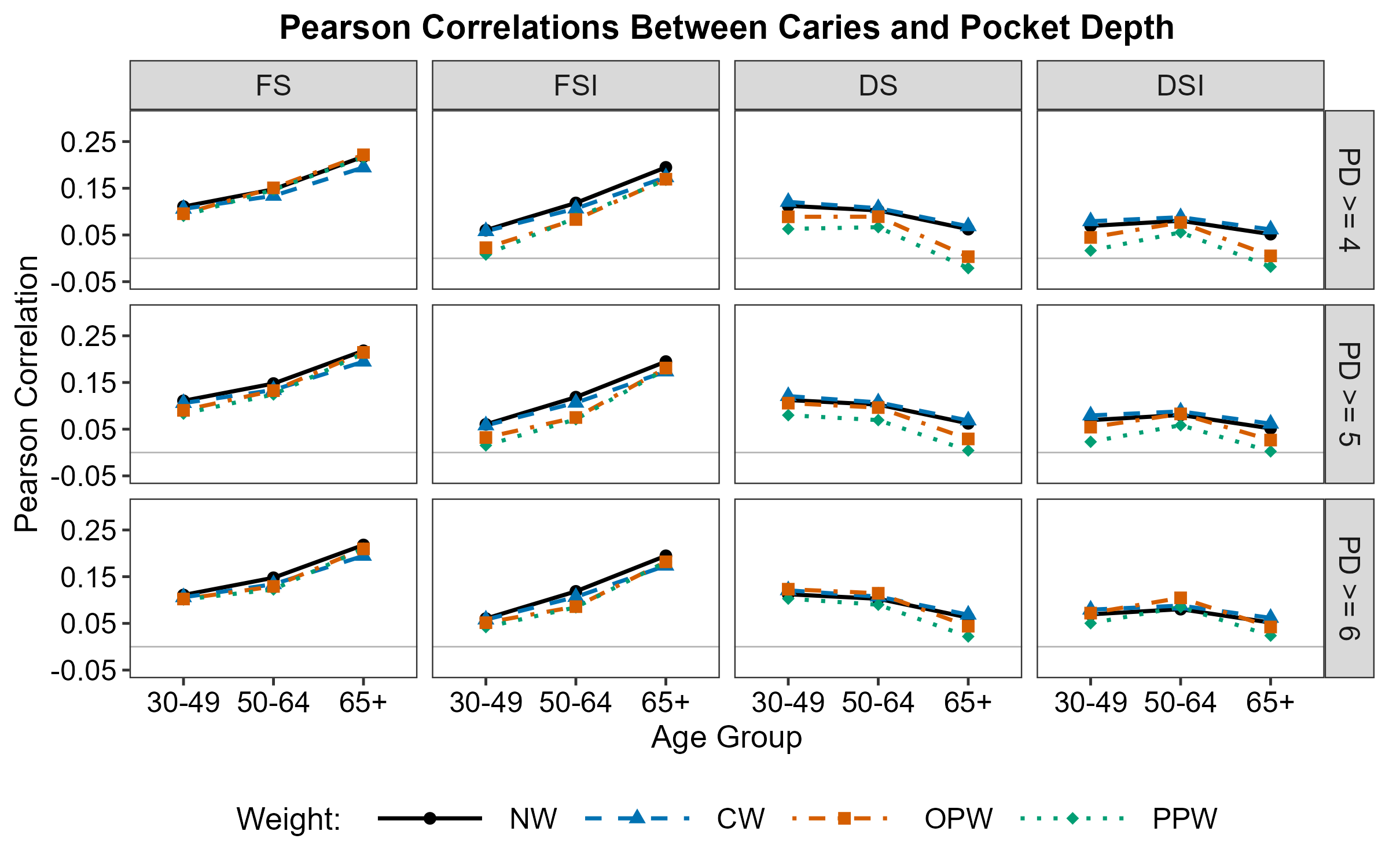}

    \captionsetup{width=0.95\linewidth}
    \caption{Pearson correlations between pocket depth (PD) and DS, DSI, FS, and FSI in NHANES across age groups and weighting methods. Rows correspond to increasing PD thresholds.}
    \label{fig:pdplots}
\end{figure}

\begin{figure}[!ht]
    \centering
    \includegraphics[width=0.75\linewidth]{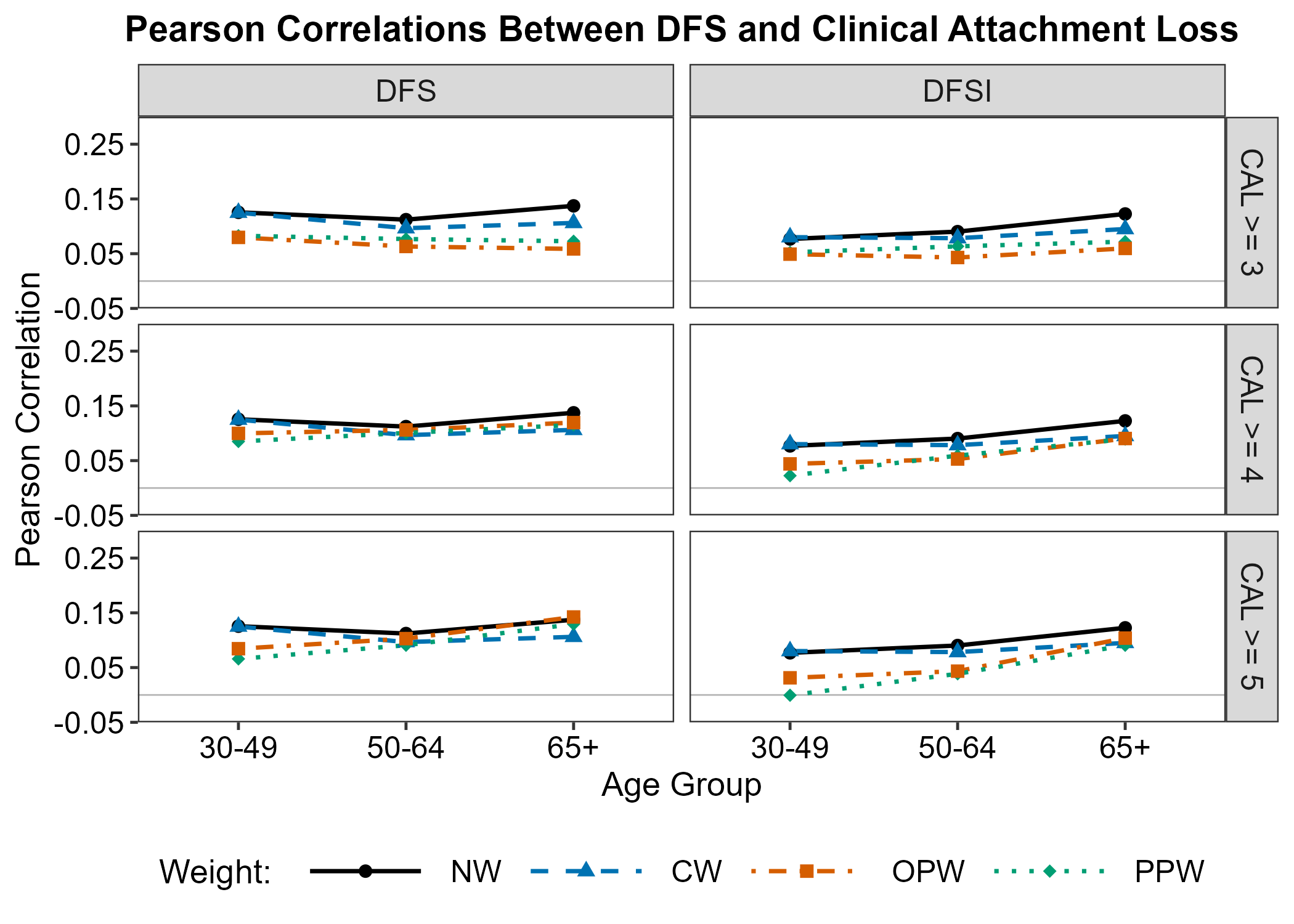}

    \captionsetup{width=0.95\linewidth}
    \caption{Pearson correlations between clinical attachment loss (CAL) and DFS and DFSI in NHANES across age groups and weighting methods. Rows correspond to increasing CAL thresholds.}
    \label{fig:dfs_cal}
\end{figure}

\begin{figure}[!ht]
    \centering
    \includegraphics[width=0.75\linewidth]{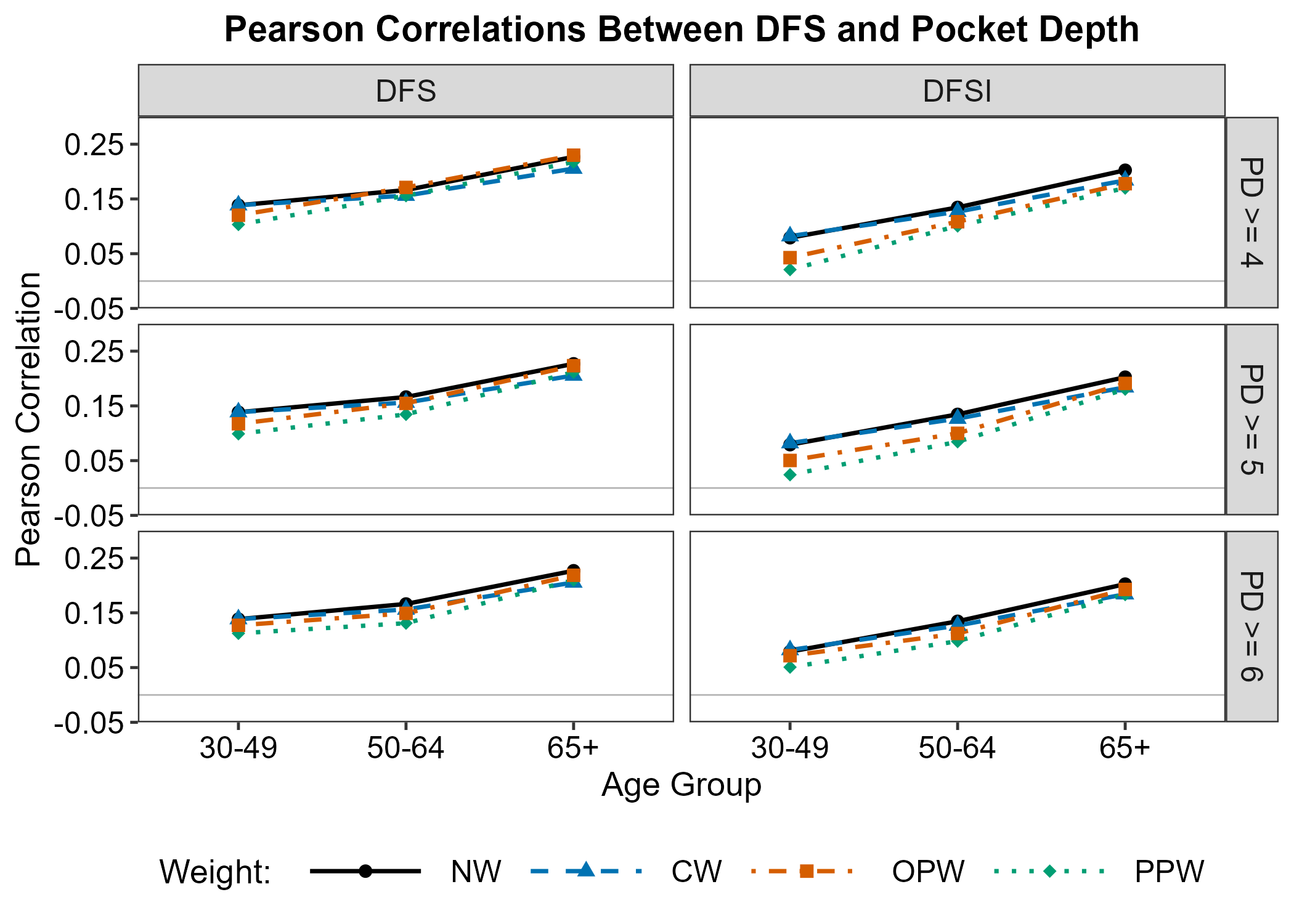}

    \captionsetup{width=0.95\linewidth}
    \caption{Pearson correlations between pocket depth (PD) and DFS and DFSI in NHANES across age groups and weighting methods. Rows correspond to increasing PD thresholds.}
    \label{fig:dfs_pd}
\end{figure}

The estimated periodontal--caries associations for PD and CAL are illustrated in Figures~\ref{fig:pdplots} and~\ref{fig:calplots}, respectively, with the corresponding DFS and DFSI associations shown in Figures~\ref{fig:dfs_pd} and~\ref{fig:dfs_cal}. Tables containing the exact values for all associations are available in the Supplementary Materials. Associations involving FS were generally modest, with correlations with CAL remaining small but positive across age groups and weighting schemes. Associations between FS and PD were more pronounced, particularly among adults aged 65 years and older. The corresponding interproximal measure, FSI, showed weaker associations with both periodontal outcomes, but with a similar tendency to increase with age.

DFS and DFSI generally followed patterns more similar to FS and FSI than to the decayed-surface measures. Associations with PD increased consistently with age, with the largest correlations observed among adults aged 65 years and older. Associations with CAL were more modest, although DFSI also tended to increase with age. For both DFS and DFSI, OPW and PPW estimates were generally attenuated relative to NW and CW, particularly among younger age groups, with the estimates becoming more similar at older ages.

Associations involving DS and DSI were generally positive and exhibited similar patterns across age groups and periodontal thresholds. For both measures, the OPW and PPW estimates were typically attenuated relative to the NW and CW estimates, which were generally similar to one another. This ordering was observed across nearly all periodontal thresholds, with exceptions for DS and DSI at $\mathrm{CAL}\geq3$ and for FS at $\mathrm{PD}\geq4$ and $\mathrm{CAL}\geq3$. The largest separation of OPW and PPW from CW and NW for FS occurred at the less severe dichotomization thresholds, whereas for DS the largest separation occurred at the more severe thresholds. This could indicate some sensitivity of the estimated association to the subgroup definition, similar to the simulation results.

Across the six caries measures, the estimated correlations were predominantly positive but modest in magnitude. DSI generally followed the same pattern as DS, whereas FSI was considerably more attenuated than FS. DFS and DFSI more closely resembled the filled-surface measures, particularly in their age-related patterns and the separation between the subgroup-adjusted and unadjusted estimates. Age-related differences were most apparent for associations involving FS, DFS, and DFSI with PD, which tended to increase with age, while associations involving DS and DSI generally decreased with age.

\FloatBarrier
\section{Discussion}\label{Discussion}

This work addresses the lack of a weighting approach for estimating marginal association between paired clustered outcomes when informative subgroup size is present. Existing marginal association measures can accommodate weighted estimating equations, so the problem is not the association measure itself, but how observations should be weighted when the subgroup sizes induced by the two outcomes are informative. Using within-cluster resampling arguments, we developed PPW and OPW to extend the usual cluster-weighting approach to this setting.

We also made a practical modification to the existing ISS testing procedure. The permutation-based test becomes computationally difficult for data containing very large numbers of clusters. Dividing the clusters into smaller independent subsets, performing the permutation test within those subsets, and combining the resulting evidence using Stouffer's method made the procedure computationally feasible without changing the underlying ISS test.

The simulations provided some guidance for how the new weights should be interpreted. When informative subgroup structure was associated primarily with the relationship between the paired observations themselves, the additional pair-based weighting could substantially reduce bias beyond CW. In contrast, when much of the association was carried through latent differences between clusters, PPW and OPW could attenuate meaningful association. We conclude that pair-based weighting should not be viewed simply as a stronger correction than cluster weighting. It changes how the within-cluster paired subgroups contribute to the association, and that distinction matters depending on where the association originates.

The categorization used to construct those subgroups is also important. PPW and OPW depend on how the outcomes are divided into clinically or scientifically meaningful groups. The heatmap analysis showed that an appropriate dichotomization could markedly improve estimation, while a poorly aligned definition could be less useful when latent cluster-level association was also present. In the setting driven primarily by within-unit association, however, PPW did not perform worse than CW for any of the dichotomizations examined. These results suggest that subgroup definitions should be chosen for their scientific interpretation rather than treated as tuning parameters.

For an ordered response with several possible dichotomizations, it is not clear that inference should depend on a single chosen cutpoint. The simulation results suggest that different cutpoints can produce meaningfully different estimates, motivating future work on procedures that incorporate information across multiple subgroup definitions rather than conditioning on one dichotomization. Such an approach could provide a more complete characterization of informative subgroup structure.

The retention mechanism was an important part of the simulation because it separated association in the underlying outcomes from association created by the observation process. Informative observation alone could produce ISS and could introduce an apparent association even when none was present in the complete data. This is particularly relevant to dental data, where the teeth available for analysis may reflect previous disease, treatment, and tooth loss. In this same regard, a limitation of the NHANES application is that the reason a tooth is missing does not identify whether the underlying disease process was specifically caries, periodontal disease, or another cause. We therefore cannot determine whether an outcome-dependent missing-not-at-random mechanism contributes to the observed associations or subgroup structure. The simulation results suggest that such a mechanism could be important, making this a natural direction for future work.

The NHANES application suggests that the periodontal--caries associations depend on the particular disease process being represented. Filled surfaces reflect accumulated caries experience and prior treatment, whereas decayed surfaces more directly represent untreated disease present at the time of examination. A similar distinction exists between CAL and PD: CAL reflects accumulated periodontal attachment loss and therefore contains information about disease history, while PD more directly characterizes current periodontal pocketing. These differences in time scale may help explain the contrasting age patterns observed across the caries and periodontal measures, with cumulative measures potentially reflecting shared disease history more strongly in older individuals. The attenuation observed for FSI relative to FS further suggests that this pattern is not simply reproduced when caries experience is restricted to interproximal surfaces.

The NHANES analysis reported marginal associations and did not jointly account for individual- or tooth-level covariates that may influence both periodontal and caries outcomes. The age-stratified analyses demonstrate that the estimated associations vary across broad age groups, but factors such as sex, smoking, fluoride exposure, diet, access to dental care, oral hygiene, tooth type, and prior treatment history may also contribute to the observed patterns. An important extension of this work is therefore to model the association measure itself as a function of such covariates. In particular, regression methods based on the estimated marginal association measures could help determine whether the weighting patterns observed here persist after accounting for demographic, behavioral, and clinical characteristics.

A limitation of the present research is that the bidirectional framework characterizes marginal informativeness separately in each direction, without determining whether the two directions arise from a common joint mechanism. To our knowledge, a formal procedure for characterizing such joint informativeness in paired clustered outcomes has not yet been developed. Along with methods that incorporate information across multiple subgroup definitions, this represents a natural direction for future methodological research.

Our recommendation to researchers is to consider PPW or OPW when estimating marginal association between paired clustered outcomes and there is evidence that the within-cluster sizes of the paired outcome subgroups are informative. These weights are most useful when the informativeness is tied to the relationship between the paired observations or to an outcome-dependent observation mechanism, because they reduce the influence of disproportionately represented paired subgroups. Additionally, the subgroup definitions used to construct the weights should be chosen on scientific or clinical grounds, and the resulting weighted estimates interpreted according to the corresponding marginal target. However, these weights should not be viewed as generally preferable to conventional cluster weighting outside of these specific settings. 

In conclusion, PPW and OPW provide approaches for incorporating informative subgroup structure into marginal association estimation. The simulations demonstrate that their performance depends on both the source of informativeness and the subgroup definition, particularly in the presence of latent cluster-level association.

\bibliographystyle{unsrtnat}
\bibliography{references}

\end{document}

\justifying
\maketitle

\section{Downloadable Materials}
All supererogatory illustrations, results, and commented code are available in summary files at the github repository: \url{http://www.github.com/owenvisser/icspack}. This repository allows for the complete reconstruction of the entire analysis.

\section{Methods}

\subsection{Resampling Justification of the Proposed Weights}

The proposed weights can be viewed as deterministic counterparts of
within-cluster resampling schemes. We give a common derivation for both
population pair weighting (PPW) and observed pair weighting (OPW). For cluster $i$, let
\begin{align}
\mathcal P_i
=
\{P_{ij}:j=1,\ldots,n_i\}
\end{align}
denote the set of paired categories observed in the cluster, and let
\begin{align}
N_{iP}=|\mathcal P_i|.
\end{align}
For any paired category $p$, define
\begin{align}
n_{ip}
=
\sum_{j=1}^{n_i} I(P_{ij}=p).
\end{align}
Thus, for unit $j$, $n_{i,P_{ij}}$ is the number of observations in
cluster $i$ having the same paired category as that unit. Let $U_{ij}(\theta)$ denote the contribution of unit $j$ in cluster $i$
to an estimating equation for parameter $\theta$. Consider a two-stage resampling scheme. First, a paired category $p$ is
selected with probability $q_i(p)$. If that category is observed in
cluster $i$, one of its $n_{ip}$ observations is selected uniformly.
If the selected category is absent from the cluster, the cluster
contributes zero for that resample. For an observed unit $j$,
\begin{align}
\Pr(J_i=j\mid V_i)
=
\frac{q_i(P_{ij})}{n_{i,P_{ij}}}.
\end{align}
Consequently, if
\begin{align}
S_i(\theta)
=
\begin{cases}
U_{iJ_i}(\theta), & J_i\neq 0,\\
0, & J_i=0,
\end{cases}
\end{align}
then
\begin{equation}
E\{S_i(\theta)\mid V_i\}
=
\sum_{j=1}^{n_i}
\frac{q_i(P_{ij})}{n_{i,P_{ij}}}
U_{ij}(\theta).
\label{eq:general_resampling}
\end{equation}
Equation \eqref{eq:general_resampling} provides both proposed weighting
schemes by changing only the first-stage distribution $q_i$.

\subsubsection{Population pair weighting}

Let $\mathcal P$ denote the fixed set of $N_P$ paired categories in the
target population. Under PPW, the first-stage category is sampled
uniformly from $\mathcal P$, so that
\begin{align}
q_i(p)=\frac{1}{N_P},
\qquad p\in\mathcal P.
\end{align}
An unobserved category therefore produces a null draw. Substituting into \eqref{eq:general_resampling},
\begin{align}
E\{S_i^{\mathrm{PPW}}(\theta)\mid V_i\}
=
\frac{1}{N_P}
\sum_{j=1}^{n_i}
\frac{1}{n_{i,P_{ij}}}
U_{ij}(\theta).
\end{align}
Summing over clusters gives a common multiplicative factor $1/N_P$,
which does not affect the root of the estimating equation. Hence PPW
corresponds to the weights
\begin{equation}
\omega_{ij}^{\mathrm{PPW}}
=
\frac{1}{n_{i,P_{ij}}}.
\label{eq:ppw_supp}
\end{equation}
Thus, every paired category represented within a cluster receives the
same total weight, while categories absent from that cluster contribute
nothing.

\subsubsection{Observed pair weighting}

Under OPW, the first-stage sampling distribution is restricted to the
categories actually observed in cluster $i$:
\begin{align}
q_i(p)=\frac{1}{N_{iP}},
\qquad p\in\mathcal P_i.
\end{align}
Equation \eqref{eq:general_resampling} therefore gives
\begin{align}
E\{S_i^{\mathrm{OPW}}(\theta)\mid V_i\}
=
\sum_{j=1}^{n_i}
\frac{1}{N_{iP}n_{i,P_{ij}}}
U_{ij}(\theta).
\end{align}
The corresponding weights are
\begin{equation}
\omega_{ij}^{\mathrm{OPW}}
=
\frac{1}{N_{iP}n_{i,P_{ij}}}.
\label{eq:opw_supp}
\end{equation}
% For every observed paired category $p\in\mathcal P_i$,
% \begin{align}
% \sum_{j:P_{ij}=p}
% \omega_{ij}^{\mathrm{OPW}}
% =
% \frac{1}{N_{iP}},
% \end{align}
% and therefore
% \begin{align}
% \sum_{j=1}^{n_i}
% \omega_{ij}^{\mathrm{OPW}}
% =
% 1.
% \end{align}
% Thus OPW gives equal total weight to each cluster and distributes that
% weight equally across the paired categories represented within it. By contrast,
% \begin{align}
% \sum_{j=1}^{n_i}
% \omega_{ij}^{\mathrm{PPW}}
% =
% N_{iP},
% \end{align}
% so PPW permits clusters containing a larger number of represented
% population pair categories to contribute more total weight.

\subsubsection{Relation to within-cluster resampling}

The preceding argument establishes an exact relationship at the level of
the estimating function: the deterministic weighted estimating equation
is the conditional expectation of the corresponding resampled estimating equation. Let $R\in\{\mathrm{PPW},\mathrm{OPW}\}$ and let
\begin{align}
S_q^R(\theta)
=
\sum_{i=1}^M S_{iq}^R(\theta)
\end{align}
denote the estimating function from resample $q$. Define
\begin{align}
G_M^R(\theta)
=
E\{S_q^R(\theta)\mid V_1,\ldots,V_M\},
\end{align}
and let $\widehat\theta_R$ solve
\begin{align}
G_M^R(\widehat\theta_R)=0.
\end{align}
Under the usual regularity conditions for estimating equations and
within-cluster resampling, a first-order expansion about the population
parameter $\theta_R$ gives
\begin{align}
\widehat\theta_q^R-\theta_R
=
-A_R^{-1}\frac{1}{M}S_q^R(\theta_R)
+
o_p(M^{-1/2}),
\end{align}
where $A_R$ is the non-singular, limiting derivative matrix. Averaging
over resamples and taking the number of resamples to infinity yields,
conditionally on the observed data,
\begin{align}
\frac{1}{Q}\sum_{q=1}^Q S_q^R(\theta_R)
\longrightarrow
G_M^R(\theta_R).
\end{align}
The deterministic weighted estimator has the corresponding expansion
\begin{align}
\widehat\theta_R-\theta_R
=
-A_R^{-1}\frac{1}{M}G_M^R(\theta_R)
+
o_p(M^{-1/2}).
\end{align}
Hence the infinite-resample WCR estimator and the deterministic weighted
estimator have the same first-order asymptotic representation. Therefore PPW and OPW may be implemented directly through their weighted
estimating equations. Standard
cluster-level sandwich inference then applies\cite{williamson2003marginal}.

\subsection{Useful Limiting Cases}

The proposed weights connect naturally to existing weighting schemes. If only one margin varies in category membership within cluster $i$,
then the paired categories reduce to the subgroups induced by the other
margin. For example, if the $X$ margin has only one category, then
\begin{align}
n_{i,P_{ij}}=n_{i,L_{ij}},
\end{align}
and
\begin{align}
\omega_{ij}^{\mathrm{PPW}}
=
\frac{1}{n_{i,L_{ij}}},
\end{align}
which recovers the weighting for a single subgroup, DWGEE1, used by Huang and Leroux (2014), up to a common constant\cite{huang2011informative}. If a cluster contains only one observed paired category, then
$N_{iP}=1$ and $n_{i,P_{ij}}=n_i$, so both methods reduce to
\begin{align}
\omega_{ij}^{\mathrm{PPW}}
=
\omega_{ij}^{\mathrm{OPW}}
=
\frac{1}{n_i},
\end{align}
the inverse cluster weight\cite{williamson2003marginal}. At the opposite extreme, if every observation occupies a distinct
paired category, then $N_{iP}=n_i$ and $n_{i,P_{ij}}=1$. Consequently,
\begin{align}
\omega_{ij}^{\mathrm{PPW}}=1,
\qquad
\omega_{ij}^{\mathrm{OPW}}=\frac{1}{n_i}.
\end{align}
Thus increasingly fine categorization causes PPW to approach ordinary
unit weighting, whereas OPW retains equal total contribution from each
cluster.

\subsection{Complete-Data Correlation in the Simulation Model}

Write the generating model as
\begin{align}
X_{ij}
=
\alpha_x+\beta_xU_i+\varepsilon_{x,ij},
\qquad
Y_{ij}
=
\alpha_y+\beta_yV_i+\varepsilon_{y,ij},
\end{align}
where
\begin{align}
\operatorname{Cov}(U_i,V_i)
=
\rho_{uv}\sigma_u\sigma_v,
\qquad
\operatorname{Cov}(\varepsilon_{x,ij},\varepsilon_{y,ij})
=
\rho_{xy}\sigma_x\sigma_y,
\end{align}
and the cluster effects are independent of the unit-level errors. It follows that
\begin{align}
\operatorname{Var}(X_{ij})
=
\beta_x^2\sigma_u^2+\sigma_x^2,
\qquad
\operatorname{Var}(Y_{ij})
=
\beta_y^2\sigma_v^2+\sigma_y^2,
\end{align}
while
\begin{align}
\operatorname{Cov}(X_{ij},Y_{ij})
=
\beta_x\beta_y\rho_{uv}\sigma_u\sigma_v
+
\rho_{xy}\sigma_x\sigma_y.
\end{align}

Therefore the complete-data marginal Pearson correlation used as the
simulation target is
\begin{equation}
\rho_0
=
\frac{
\beta_x\beta_y\rho_{uv}\sigma_u\sigma_v
+
\rho_{xy}\sigma_x\sigma_y
}{
\sqrt{\beta_x^2\sigma_u^2+\sigma_x^2}
\sqrt{\beta_y^2\sigma_v^2+\sigma_y^2}
}.
\label{eq:true_corr_supp}
\end{equation}

\subsection{Combination of Subset-Level ISS Tests}

Let $\mathcal S_g$, $g=1,\ldots,G$, denote disjoint subsets of clusters, and let $p_g$ be the p-value obtained by applying the ISS test within subset $\mathcal S_g$. Because the subsets contain non-overlapping independent clusters, the subset-level test statistics are independent under the global null. For continuous null p-values,
\begin{align}
Z_g=\Phi^{-1}(1-p_g)
\end{align}
is standard normal. Stouffer's statistic is therefore
\begin{align}
Z_{\mathrm{Stouffer}}
=
\frac{1}{\sqrt G}
\sum_{g=1}^G Z_g,
\end{align}
which is standard normal under the global null, giving the combined
one-sided p-value
\begin{align}
p_{\mathrm{Stouffer}}
=
1-\Phi(Z_{\mathrm{Stouffer}}).
\end{align}
Permutation p-values are discrete rather than exactly continuous
Uniform$(0,1)$ variables, so the standard-normal calibration should be
viewed as the usual Stouffer approximation when finite permutation
distributions are used. In implementation, values numerically equal to
zero or one were truncated before applying $\Phi^{-1}$.

\newpage
\section{NHANES Association Tables}

% ============================================================
% DS
% ============================================================

\begin{table}[!ht]
\centering
\small
\setlength{\tabcolsep}{7pt}

\begin{tabular*}{0.95\linewidth}{@{\extracolsep{\fill}}lll rrr}
\toprule
\multicolumn{3}{l}{Association} & \multicolumn{3}{l}{Age Group} \\
\cmidrule{1-3} \cmidrule{4-6}
Caries & Periodontal & Weight & 30--49 & 50--64 & 65+ \\
\midrule

% DS vs CAL
DS & CAL & NW & $0.125\,(0.011)$ & $0.112\,(0.011)$ & $0.080\,(0.013)$ \\
\addlinespace[4pt]
& CAL & CW & $0.135\,(0.014)$ & $0.116\,(0.014)$ & $0.094\,(0.016)$ \\
\addlinespace[4pt]
& CAL $\geq$ 3mm & PPW & $0.099\,(0.012)$ & $0.132\,(0.015)$ & $0.084\,(0.021)$ \\
& CAL $\geq$ 4mm & -- & $0.074\,(0.012)$ & $0.082\,(0.015)$ & $0.025\,(0.018)$ \\
& CAL $\geq$ 5mm & -- & $0.075\,(0.013)$ & $0.076\,(0.018)$ & $0.006\,(0.019)$ \\
\addlinespace[4pt]
& CAL $\geq$ 3mm & OPW & $0.121\,(0.013)$ & $0.138\,(0.016)$ & $0.103\,(0.021)$ \\
& CAL $\geq$ 4mm & -- & $0.104\,(0.012)$ & $0.103\,(0.015)$ & $0.054\,(0.019)$ \\
& CAL $\geq$ 5mm & -- & $0.109\,(0.012)$ & $0.097\,(0.015)$ & $0.040\,(0.017)$ \\
\midrule

% DS vs PD
DS & PD & NW & $0.112\,(0.010)$ & $0.103\,(0.009)$ & $0.062\,(0.011)$ \\
\addlinespace[4pt]
& PD & CW & $0.121\,(0.013)$ & $0.107\,(0.011)$ & $0.069\,(0.013)$ \\
\addlinespace[4pt]
& PD $\geq$ 4mm & PPW & $0.063\,(0.012)$ & $0.067\,(0.015)$ & $-0.021\,(0.016)$ \\
& PD $\geq$ 5mm & -- & $0.080\,(0.014)$ & $0.069\,(0.015)$ & $0.004\,(0.017)$ \\
& PD $\geq$ 6mm & -- & $0.103\,(0.015)$ & $0.090\,(0.016)$ & $0.022\,(0.019)$ \\
\addlinespace[4pt]
& PD $\geq$ 4mm & OPW & $0.089\,(0.012)$ & $0.089\,(0.013)$ & $0.004\,(0.013)$ \\
& PD $\geq$ 5mm & -- & $0.106\,(0.013)$ & $0.096\,(0.013)$ & $0.029\,(0.014)$ \\
& PD $\geq$ 6mm & -- & $0.123\,(0.013)$ & $0.114\,(0.014)$ & $0.044\,(0.015)$ \\

\bottomrule
\end{tabular*}

\captionsetup{width=0.95\linewidth}
\caption{Tooth-level Pearson correlations between DS and periodontal outcomes by weighting method and age group. Standard errors are reported in parentheses. NW denotes the use of NHANES sampling weights without additional informative cluster-size weighting. PPW and OPW estimates are shown for alternative dichotomizations of the periodontal outcome.}
\label{tab:nhanes_DS}
\end{table}

% ============================================================
% DSI
% ============================================================

\begin{table}[!ht]
\centering
\small
\setlength{\tabcolsep}{7pt}

\begin{tabular*}{0.95\linewidth}{@{\extracolsep{\fill}}lll rrr}
\toprule
\multicolumn{3}{l}{Association} & \multicolumn{3}{l}{Age Group} \\
\cmidrule{1-3} \cmidrule{4-6}
Caries & Periodontal & Weight & 30--49 & 50--64 & 65+ \\
\midrule

% DSI vs CAL
DSI & CAL & NW & $0.086\,(0.011)$ & $0.097\,(0.011)$ & $0.075\,(0.014)$ \\
\addlinespace[4pt]
& CAL & CW & $0.098\,(0.014)$ & $0.105\,(0.013)$ & $0.095\,(0.018)$ \\
\addlinespace[4pt]
& CAL $\geq$ 3mm & PPW & $0.081\,(0.015)$ & $0.142\,(0.018)$ & $0.104\,(0.025)$ \\
& CAL $\geq$ 4mm & -- & $0.044\,(0.014)$ & $0.086\,(0.017)$ & $0.037\,(0.024)$ \\
& CAL $\geq$ 5mm & -- & $0.040\,(0.015)$ & $0.077\,(0.018)$ & $0.018\,(0.023)$ \\
\addlinespace[4pt]
& CAL $\geq$ 3mm & OPW & $0.098\,(0.015)$ & $0.142\,(0.018)$ & $0.120\,(0.026)$ \\
& CAL $\geq$ 4mm & -- & $0.072\,(0.013)$ & $0.101\,(0.017)$ & $0.061\,(0.024)$ \\
& CAL $\geq$ 5mm & -- & $0.073\,(0.013)$ & $0.094\,(0.015)$ & $0.045\,(0.020)$ \\
\midrule

% DSI vs PD
DSI & PD & NW & $0.069\,(0.010)$ & $0.080\,(0.009)$ & $0.052\,(0.011)$ \\
\addlinespace[4pt]
& PD & CW & $0.079\,(0.012)$ & $0.088\,(0.012)$ & $0.062\,(0.013)$ \\
\addlinespace[4pt]
& PD $\geq$ 4mm & PPW & $0.017\,(0.013)$ & $0.056\,(0.017)$ & $-0.018\,(0.021)$ \\
& PD $\geq$ 5mm & -- & $0.023\,(0.014)$ & $0.058\,(0.017)$ & $0.002\,(0.021)$ \\
& PD $\geq$ 6mm & -- & $0.050\,(0.015)$ & $0.085\,(0.019)$ & $0.024\,(0.022)$ \\
\addlinespace[4pt]
& PD $\geq$ 4mm & OPW & $0.045\,(0.011)$ & $0.077\,(0.016)$ & $0.005\,(0.016)$ \\
& PD $\geq$ 5mm & -- & $0.054\,(0.011)$ & $0.083\,(0.014)$ & $0.027\,(0.015)$ \\
& PD $\geq$ 6mm & -- & $0.072\,(0.012)$ & $0.105\,(0.015)$ & $0.042\,(0.016)$ \\

\bottomrule
\end{tabular*}

\captionsetup{width=0.95\linewidth}
\caption{Tooth-level Pearson correlations between DSI and periodontal outcomes by weighting method and age group. Standard errors are reported in parentheses. NW denotes the use of NHANES sampling weights without additional informative cluster-size weighting. PPW and OPW estimates are shown for alternative dichotomizations of the periodontal outcome.}
\label{tab:nhanes_DSI}
\end{table}

% ============================================================
% FS
% ============================================================

\begin{table}[!ht]
\centering
\small
\setlength{\tabcolsep}{7pt}

\begin{tabular*}{0.95\linewidth}{@{\extracolsep{\fill}}lll rrr}
\toprule
\multicolumn{3}{l}{Association} & \multicolumn{3}{l}{Age Group} \\
\cmidrule{1-3} \cmidrule{4-6}
Caries & Periodontal & Weight & 30--49 & 50--64 & 65+ \\
\midrule

% FS vs CAL
FS & CAL & NW & $0.094\,(0.008)$ & $0.092\,(0.012)$ & $0.126\,(0.013)$ \\
\addlinespace[4pt]
& CAL & CW & $0.088\,(0.008)$ & $0.072\,(0.012)$ & $0.092\,(0.015)$ \\
\addlinespace[4pt]
& CAL $\geq$ 3mm & PPW & $0.081\,(0.007)$ & $0.073\,(0.008)$ & $0.078\,(0.008)$ \\
& CAL $\geq$ 4mm & -- & $0.075\,(0.009)$ & $0.092\,(0.010)$ & $0.119\,(0.011)$ \\
& CAL $\geq$ 5mm & -- & $0.048\,(0.010)$ & $0.078\,(0.012)$ & $0.131\,(0.014)$ \\
\addlinespace[4pt]
& CAL $\geq$ 3mm & OPW & $0.066\,(0.009)$ & $0.046\,(0.009)$ & $0.056\,(0.010)$ \\
& CAL $\geq$ 4mm & -- & $0.078\,(0.009)$ & $0.084\,(0.011)$ & $0.115\,(0.012)$ \\
& CAL $\geq$ 5mm & -- & $0.055\,(0.009)$ & $0.078\,(0.012)$ & $0.134\,(0.014)$ \\
\midrule

% FS vs PD
FS & PD & NW & $0.111\,(0.008)$ & $0.148\,(0.012)$ & $0.218\,(0.013)$ \\
\addlinespace[4pt]
& PD & CW & $0.106\,(0.008)$ & $0.134\,(0.013)$ & $0.195\,(0.014)$ \\
\addlinespace[4pt]
& PD $\geq$ 4mm & PPW & $0.091\,(0.010)$ & $0.147\,(0.013)$ & $0.216\,(0.015)$ \\
& PD $\geq$ 5mm & -- & $0.083\,(0.011)$ & $0.124\,(0.016)$ & $0.212\,(0.018)$ \\
& PD $\geq$ 6mm & -- & $0.101\,(0.011)$ & $0.123\,(0.017)$ & $0.210\,(0.019)$ \\
\addlinespace[4pt]
& PD $\geq$ 4mm & OPW & $0.096\,(0.010)$ & $0.151\,(0.013)$ & $0.222\,(0.015)$ \\
& PD $\geq$ 5mm & -- & $0.090\,(0.010)$ & $0.133\,(0.014)$ & $0.214\,(0.016)$ \\
& PD $\geq$ 6mm & -- & $0.102\,(0.010)$ & $0.129\,(0.014)$ & $0.209\,(0.016)$ \\

\bottomrule
\end{tabular*}

\captionsetup{width=0.95\linewidth}
\caption{Tooth-level Pearson correlations between FS and periodontal outcomes by weighting method and age group. Standard errors are reported in parentheses. NW denotes the use of NHANES sampling weights without additional informative cluster-size weighting. PPW and OPW estimates are shown for alternative dichotomizations of the periodontal outcome.}
\label{tab:nhanes_FS}
\end{table}

% ============================================================
% FSI
% ============================================================

\begin{table}[!ht]
\centering
\small
\setlength{\tabcolsep}{7pt}

\begin{tabular*}{0.95\linewidth}{@{\extracolsep{\fill}}lll rrr}
\toprule
\multicolumn{3}{l}{Association} & \multicolumn{3}{l}{Age Group} \\
\cmidrule{1-3} \cmidrule{4-6}
Caries & Periodontal & Weight & 30--49 & 50--64 & 65+ \\
\midrule

% FSI vs CAL
FSI & CAL & NW & $0.053\,(0.007)$ & $0.070\,(0.012)$ & $0.111\,(0.013)$ \\
\addlinespace[4pt]
& CAL & CW & $0.049\,(0.007)$ & $0.053\,(0.012)$ & $0.078\,(0.014)$ \\
\addlinespace[4pt]
& CAL $\geq$ 3mm & PPW & $0.044\,(0.008)$ & $0.054\,(0.008)$ & $0.073\,(0.008)$ \\
& CAL $\geq$ 4mm & -- & $0.010\,(0.010)$ & $0.045\,(0.010)$ & $0.089\,(0.011)$ \\
& CAL $\geq$ 5mm & -- & $-0.017\,(0.012)$ & $0.020\,(0.012)$ & $0.088\,(0.013)$ \\
\addlinespace[4pt]
& CAL $\geq$ 3mm & OPW & $0.027\,(0.010)$ & $0.014\,(0.010)$ & $0.048\,(0.010)$ \\
& CAL $\geq$ 4mm & -- & $0.020\,(0.010)$ & $0.024\,(0.012)$ & $0.080\,(0.012)$ \\
& CAL $\geq$ 5mm & -- & $0.006\,(0.010)$ & $0.012\,(0.013)$ & $0.091\,(0.015)$ \\
\midrule

% FSI vs PD
FSI & PD & NW & $0.060\,(0.007)$ & $0.119\,(0.012)$ & $0.195\,(0.013)$ \\
\addlinespace[4pt]
& PD & CW & $0.058\,(0.008)$ & $0.107\,(0.013)$ & $0.174\,(0.014)$ \\
\addlinespace[4pt]
& PD $\geq$ 4mm & PPW & $0.008\,(0.011)$ & $0.087\,(0.012)$ & $0.168\,(0.015)$ \\
& PD $\geq$ 5mm & -- & $0.015\,(0.012)$ & $0.071\,(0.015)$ & $0.178\,(0.017)$ \\
& PD $\geq$ 6mm & -- & $0.042\,(0.012)$ & $0.084\,(0.016)$ & $0.182\,(0.019)$ \\
\addlinespace[4pt]
& PD $\geq$ 4mm & OPW & $0.023\,(0.010)$ & $0.083\,(0.013)$ & $0.169\,(0.015)$ \\
& PD $\geq$ 5mm & -- & $0.032\,(0.011)$ & $0.075\,(0.014)$ & $0.181\,(0.016)$ \\
& PD $\geq$ 6mm & -- & $0.052\,(0.011)$ & $0.086\,(0.014)$ & $0.182\,(0.016)$ \\

\bottomrule
\end{tabular*}

\captionsetup{width=0.95\linewidth}
\caption{Tooth-level Pearson correlations between FSI and periodontal outcomes by weighting method and age group. Standard errors are reported in parentheses. NW denotes the use of NHANES sampling weights without additional informative cluster-size weighting. PPW and OPW estimates are shown for alternative dichotomizations of the periodontal outcome.}
\label{tab:nhanes_FSI}
\end{table}

% ============================================================
% DFS
% ============================================================

\begin{table}[!ht]
\centering
\small
\setlength{\tabcolsep}{7pt}

\begin{tabular*}{0.95\linewidth}{@{\extracolsep{\fill}}lll rrr}
\toprule
\multicolumn{3}{l}{Association} & \multicolumn{3}{l}{Age Group} \\
\cmidrule{1-3} \cmidrule{4-6}
Caries & Periodontal & Weight & 30--49 & 50--64 & 65+ \\
\midrule

% DFS vs CAL
DFS & CAL & NW & $0.126\,(0.008)$ & $0.112\,(0.012)$ & $0.137\,(0.013)$ \\
\addlinespace[4pt]
& CAL & CW & $0.125\,(0.008)$ & $0.097\,(0.012)$ & $0.106\,(0.014)$ \\
\addlinespace[4pt]
& CAL $\geq$ 3mm & PPW & $0.083\,(0.005)$ & $0.077\,(0.008)$ & $0.073\,(0.009)$ \\
& CAL $\geq$ 4mm & -- & $0.085\,(0.008)$ & $0.101\,(0.010)$ & $0.115\,(0.011)$ \\
& CAL $\geq$ 5mm & -- & $0.066\,(0.009)$ & $0.091\,(0.012)$ & $0.129\,(0.014)$ \\
\addlinespace[4pt]
& CAL $\geq$ 3mm & OPW & $0.080\,(0.006)$ & $0.063\,(0.009)$ & $0.059\,(0.010)$ \\
& CAL $\geq$ 4mm & -- & $0.100\,(0.008)$ & $0.106\,(0.011)$ & $0.119\,(0.012)$ \\
& CAL $\geq$ 5mm & -- & $0.084\,(0.008)$ & $0.103\,(0.012)$ & $0.142\,(0.014)$ \\
\midrule

% DFS vs PD
DFS & PD & NW & $0.138\,(0.008)$ & $0.166\,(0.012)$ & $0.227\,(0.012)$ \\
\addlinespace[4pt]
& PD & CW & $0.139\,(0.008)$ & $0.156\,(0.012)$ & $0.205\,(0.013)$ \\
\addlinespace[4pt]
& PD $\geq$ 4mm & PPW & $0.103\,(0.009)$ & $0.156\,(0.013)$ & $0.217\,(0.016)$ \\
& PD $\geq$ 5mm & -- & $0.099\,(0.010)$ & $0.134\,(0.015)$ & $0.212\,(0.019)$ \\
& PD $\geq$ 6mm & -- & $0.113\,(0.010)$ & $0.131\,(0.016)$ & $0.210\,(0.020)$ \\
\addlinespace[4pt]
& PD $\geq$ 4mm & OPW & $0.120\,(0.009)$ & $0.171\,(0.012)$ & $0.230\,(0.015)$ \\
& PD $\geq$ 5mm & -- & $0.117\,(0.009)$ & $0.155\,(0.014)$ & $0.223\,(0.016)$ \\
& PD $\geq$ 6mm & -- & $0.127\,(0.009)$ & $0.149\,(0.014)$ & $0.219\,(0.016)$ \\

\bottomrule
\end{tabular*}

\captionsetup{width=0.95\linewidth}
\caption{Tooth-level Pearson correlations between DFS and periodontal outcomes by weighting method and age group. Standard errors are reported in parentheses. NW denotes the use of NHANES sampling weights without additional informative cluster-size weighting. PPW and OPW estimates are shown for alternative dichotomizations of the periodontal outcome.}
\label{tab:nhanes_DFS}
\end{table}

% ============================================================
% DFSI
% ============================================================

\begin{table}[!ht]
\centering
\small
\setlength{\tabcolsep}{7pt}

\begin{tabular*}{0.95\linewidth}{@{\extracolsep{\fill}}lll rrr}
\toprule
\multicolumn{3}{l}{Association} & \multicolumn{3}{l}{Age Group} \\
\cmidrule{1-3} \cmidrule{4-6}
Caries & Periodontal & Weight & 30--49 & 50--64 & 65+ \\
\midrule

% DFSI vs CAL
DFSI & CAL & NW & $0.077\,(0.008)$ & $0.090\,(0.012)$ & $0.123\,(0.013)$ \\
\addlinespace[4pt]
& CAL & CW & $0.080\,(0.008)$ & $0.078\,(0.012)$ & $0.095\,(0.013)$ \\
\addlinespace[4pt]
& CAL $\geq$ 3mm & PPW & $0.052\,(0.007)$ & $0.063\,(0.007)$ & $0.072\,(0.008)$ \\
& CAL $\geq$ 4mm & -- & $0.023\,(0.009)$ & $0.059\,(0.009)$ & $0.089\,(0.011)$ \\
& CAL $\geq$ 5mm & -- & $-0.001\,(0.011)$ & $0.039\,(0.011)$ & $0.091\,(0.013)$ \\
\addlinespace[4pt]
& CAL $\geq$ 3mm & OPW & $0.049\,(0.008)$ & $0.043\,(0.010)$ & $0.060\,(0.009)$ \\
& CAL $\geq$ 4mm & -- & $0.044\,(0.009)$ & $0.053\,(0.011)$ & $0.090\,(0.012)$ \\
& CAL $\geq$ 5mm & -- & $0.031\,(0.010)$ & $0.043\,(0.012)$ & $0.104\,(0.014)$ \\
\midrule

% DFSI vs PD
DFSI & PD & NW & $0.079\,(0.008)$ & $0.135\,(0.012)$ & $0.202\,(0.013)$ \\
\addlinespace[4pt]
& PD & CW & $0.082\,(0.008)$ & $0.127\,(0.012)$ & $0.184\,(0.013)$ \\
\addlinespace[4pt]
& PD $\geq$ 4mm & PPW & $0.021\,(0.010)$ & $0.101\,(0.012)$ & $0.170\,(0.015)$ \\
& PD $\geq$ 5mm & -- & $0.024\,(0.011)$ & $0.084\,(0.014)$ & $0.180\,(0.017)$ \\
& PD $\geq$ 6mm & -- & $0.051\,(0.011)$ & $0.098\,(0.015)$ & $0.183\,(0.019)$ \\
\addlinespace[4pt]
& PD $\geq$ 4mm & OPW & $0.043\,(0.010)$ & $0.108\,(0.012)$ & $0.178\,(0.015)$ \\
& PD $\geq$ 5mm & -- & $0.050\,(0.010)$ & $0.100\,(0.014)$ & $0.191\,(0.016)$ \\
& PD $\geq$ 6mm & -- & $0.072\,(0.010)$ & $0.112\,(0.014)$ & $0.193\,(0.016)$ \\

\bottomrule
\end{tabular*}

\captionsetup{width=0.95\linewidth}
\caption{Tooth-level Pearson correlations between DFSI and periodontal outcomes by weighting method and age group. Standard errors are reported in parentheses. NW denotes the use of NHANES sampling weights without additional informative cluster-size weighting. PPW and OPW estimates are shown for alternative dichotomizations of the periodontal outcome.}
\label{tab:nhanes_DFSI}
\end{table}

\FloatBarrier

\bibliographystyle{unsrtnat}
\bibliography{references}